\documentclass[fleqn,10pt%
]{wlscirep}
\usepackage[utf8]{inputenc}
\usepackage[T1]{fontenc}
\usepackage{lineno}
\usepackage{nameref,zref-xr}
\usepackage{xr}
\usepackage{chemformula}
\usepackage{pdfpages}

\makeatletter
\newcommand*{\addFileDependency}[1]{% argument=file name and extension
  \typeout{(#1)}
  \@addtofilelist{#1}
  \IfFileExists{#1}{}{\typeout{No file #1.}}
}
\makeatother

\title {Making atomistic materials calculations accessible with the AiiDAlab Quantum ESPRESSO app}
\author[1,2,$\dag$]{Xing Wang}
\author[1,2,$\dag$]{Edan Bainglass}
\author[1,2,$\dag$]{Miki Bonacci}
\author[3,$\dag$]{Andres Ortega-Guerrero}
\author[4]{Lorenzo Bastonero}
\author[1,2]{Marnik Bercx}
\author[6,7]{Pietro Bonf\`a}
\author[8]{Roberto De Renzi}
\author[5]{Dou Du}
\author[7]{Peter N. O. Gillespie}
\author[6,7]{Michael A. Hern\'andez-Bertr\'an}
\author[9]{Daniel Hollas}
\author[5]{Sebastiaan P. Huber}
\author[6,7]{Elisa Molinari}
\author[8]{Ifeanyi J. Onuorah}
\author[1,2]{Nataliya Paulish}
\author[7]{Deborah Prezzi}
\author[5]{Junfeng Qiao}
\author[1,2]{Timo Reents}
\author[5]{Christopher J. Sewell}
\author[1,2]{Iurii Timrov}
\author[3]{Aliaksandr V. Yakutovich}
\author[1,2]{Jusong Yu}
\author[1,2,4,5]{Nicola Marzari}
\author[3,*]{Carlo A. Pignedoli}
\author[1,2,3,*]{Giovanni Pizzi}

\affil[1]{PSI Center for Scientific Computing, Theory and Data, 5232 Villigen PSI, Switzerland}
\affil[2]{National Centre for Computational Design and Discovery of Novel Materials (MARVEL), 5232 Villigen PSI, Switzerland}
\affil[3]{nanotech@surfaces Laboratory, Empa-Swiss Federal Laboratories for Materials Science and Technology, 8600 D\"ubendorf, Switzerland}
\affil[4]{U Bremen Excellence Chair, Bremen Centre for Computational Materials Science, and MAPEX Center for Materials and Processes, University of Bremen, 28359 Bremen, Germany}
\affil[5]{Theory and Simulation of Materials (THEOS), \'Ecole Polytechnique F\'ed\'erale de Lausanne, 1015 Lausanne, Switzerland}
\affil[6]{Dipartimento di Scienze Fisiche, Informatiche, Matematiche (FIM),
Universit\`a di Modena e Reggio Emilia, 41125 Modena, Italy}
\affil[7]{Nanoscience Institute, National Research Council (CNR-NANO), 41125 Modena, Italy}
\affil[8]{Department of Physics and Earth Sciences, University of Parma, 43124 Parma, Italy}
\affil[9]{Center for Computational Chemistry, School of Chemistry, University of Bristol, BS8 1TS Bristol, UK}

\affil[$\dag$]{these authors contributed equally to this work}
\affil[*]{Corresponding authors: Carlo A. Pignedoli (carlo.pignedoli@empa.ch), Giovanni Pizzi (giovanni.pizzi@psi.ch)}

\begin{abstract}
Despite the wide availability of density functional theory (DFT) codes, their adoption by the broader materials science community remains limited due to challenges such as software installation, input preparation, high-performance computing setup, and output analysis. To overcome these barriers, we introduce the \textsc{Quantum ESPRESSO} app, an intuitive, web-based platform built on AiiDAlab that integrates user-friendly graphical interfaces with automated DFT workflows. The app employs a modular Input-Process-Output model and a plugin-based architecture, providing predefined computational protocols, automated error handling, and interactive results visualization. We demonstrate the app's capabilities through plugins for electronic band structures, projected density of states, phonon, infrared/Raman, X-ray and muon spectroscopies, Hubbard parameters (DFT+$U$+$V$), Wannier functions, and post-processing tools. By extending the FAIR principles to simulations, workflows, and analyses, the app enhances the accessibility and reproducibility of advanced DFT calculations and provides a general template to interface with other first-principles calculation codes.

\end{abstract}
\begin{document}

\flushbottom
\maketitle

\thispagestyle{empty}

\section*{Introduction}

Density-functional theory (DFT) simulations play a pivotal role in condensed matter physics and chemistry by providing insights and predictions for electronic structure and properties of materials at the atomistic level~\cite{hohenberg1964inhomogeneous, kohn1965self}. These simulations enable researchers to predict a wide range of properties---such as electronic, magnetic, optical, mechanical, and thermal behavior---aiding in the design of novel materials for applications across diverse fields, including semiconductors, catalysts, energy storage systems, quantum materials, nanotechnology, and more~\cite{marzari2021electronic}. 

A multitude of powerful and widely used simulation codes have been developed in the past decades~\cite{Talirz_Ghiringhelli_Smit_2021} and cross-verified~\cite{lejaeghere2016reproducibility,bosoni2024verify} to perform these calculations precisely and efficiently.
Despite these advances, performing DFT calculations remains a complex, multi-stage process that demands significant time and effort, particularly for non-specialists. By non-specialists, we refer to researchers whose primary expertise lies outside computational materials science, such as experimentalists, industrial R\&D engineers, educators, and students. These users often face steep learning curves due to obstacles including: (1) software installation and configuration, (2) code-specific input file preparation, (3) parameter tuning, (4) results interpretation, (5) setup and utilization of high-performance computing (HPC) resources, and (6) validation and benchmarking against experimental data or established theoretical values. Furthermore, many property calculations require orchestrating long sequences of interdependent simulations---sometimes numbering in the tens or even hundreds. For instance, simulating phonon properties requires performing total energy calculations for each unique finite displacement (in supercells) as well as, if long-range dipole corrections are needed, for each applied finite electric field (in the unit cell). Even seasoned computational materials scientists (whom we will refer to as specialists in the following) routinely encounter such time-consuming, error-prone workflows that distract them from scientific insight.

Consequently, there is a growing demand for platforms that simultaneously (1) lower the entry barrier for non-specialists, enabling them to perform reliable simulations with minimal friction (for example, making advanced workflows accessible to experimentalists to help them interpret experimental results or guide the design of new experiments); and (2) provide specialists with robust, reproducible, and extensible infrastructures and advanced interactive analysis to enhance productivity in their simulations, as well as to allow them to incorporate their domain expertise into the platform, making it accessible to non-specialists.
Achieving this goal requires a combination of: (1) ensuring that simulation codes use optimal computational settings with minimal user intervention;
(2) implementing workflows that can detect and recover from common failures (e.g., switching to more robust algorithms in cases of convergence issues); (3) providing user-friendly user interfaces (UIs) to assist non-specialists in preparing inputs, submitting and monitoring workflows, and analyzing results; and (4) enabling easy sharing and reusing of data to foster transparency and collaboration.

Over the past decade, formal workflows have become increasingly important in computational materials science, enabling the automation of simulations and supporting large-scale, high-throughput studies that generate extensive databases of material properties~\cite{blum2024roadmap, MaterialsProject_2013, curtarolo_high-throughput_2013, cheon_data_2017, Mounet_2018,MC2DB, vecchio_high-throughput_2021, Gjerding_2021, MC3D}. Available workflow frameworks include AFLOW~\cite{curtarolo2012aflow}, AiiDA~\cite{PIZZI2016, UHRIN2021, huber2020aiida}, ASR+ASE+myqueue~\cite{gjerding2021atomic, mortensen2020myqueue}, Atomate+FireWorks~\cite{mathew2017atomate, jain2015fireworks}, Covalent~\cite{will_cunningham_2025_15400489}, Jobflow~\cite{rosen2024jobflow}, HTTK~\cite{armiento2020database}, MISPR~\cite{atwi2022mispr}, Pyiron~\cite{janssen2019pyiron}, and QMPY~\cite{kirklin2015open}, among others. These frameworks facilitate researchers in managing automated sequences of simulations, and often promote the FAIR (Findable, Accessible, Interoperable, and Reusable) data principles~\cite{wilkinson2016fair}, making it easier to share and reuse data generated by workflows among researchers, thus advancing the collective understanding of material behavior. One noteworthy advancement in this field has been the development of common workflows for computing material properties across different quantum engines~\cite{huber2021common,gjerding2021atomic,ganose2025_atomate2,quacc2024}. These workflows provide a standardized approach to property calculations, improving reproducibility and facilitating direct comparison and verification between different DFT implementations~\cite{bosoni2024verify}. Despite these advancements, existing platforms often require users to formulate workflows through scripting, necessitating programming skills that many non-specialists may not master. Additionally, these frameworks do not address the practical challenges such as software installation and computational resources setup. Therefore, despite the advantages of workflows, significant barriers remain for non-specialists wishing to leverage these tools.

Graphical user interfaces (GUIs) could remedy this situation. However, developing GUIs for workflows traditionally requires in-depth knowledge of both workflow engines and graphical programming languages. While commercial offerings such as Materials Studio (https://www.3ds.com/products/biovia/materials-studio), MedeA (https://www.materialsdesign.com/medea-software), SimStack~\cite{rego2022simstack}, ASAP (https://www.simuneatomistics.com/asap/), Mat3ra (https://www.mat3ra.com/) and Materys (https://www.materys.com/) provide GUIs for setting up and running simulations, they are often proprietary and with limited capability for quick adaption or user extension to custom workflows. On the other hand, some open-source computational codes do offer their own GUIs to facilitate simulation setup and analysis, such as GPAW~\cite{mortensen2024gpaw} and PWgui (http://www-k3.ijs.si/kokalj/pwgui/pwgui.html ). However, these GUIs are typically limited to running single calculations for specific codes, making it difficult to switch between different software or support higher-level workflows. Other frameworks---such as CINEMAS~\cite{gupta2021cinemas} and ALKEMIE~\cite{wang2021alkemie}---have emerged, and these provide user-friendly GUIs and integrated environments for advanced materials simulations. However, these solutions generally require a local installation, highlighting the ongoing need for more flexible, web-based platforms that can accommodate both local and cloud-based usage while remaining widely accessible to the research community.

In our previous work, we developed AiiDAlab~\cite{YAKUTOVICH2021110165}, a web-based platform that enables computational scientists to package scientific workflows and computational environments into user-friendly applications (apps). AiiDAlab integrates AiiDA workflows with graphical interfaces (based on Jupyter notebooks) for execution and data analysis, making them accessible to a broader audience. In this work, leveraging the AiiDAlab platform, we identify a common structure when designing an app and GUI for DFT calculations, and develop a generic system that can serve as a template for other simulation codes. We demonstrate this concept by implementing a feature-rich app specifically designed for \textsc{Quantum ESPRESSO} (QE)~\cite{giannozzi2009,giannozzi2017,giannozzi2020quantum}, an open-source and widely used DFT simulation package~\cite{Talirz_Ghiringhelli_Smit_2021}. In the following sections, we describe the architecture and general structure of the app, presenting some implementation details including the integration with AiiDA and AiiDAlab.
We then demonstrate how it addresses the challenges faced by non-specialists users in performing DFT simulations by showcasing the app's features through example use cases. 
These features leverage several AiiDA workflows recently developed for computing relevant materials properties (ranging from electronic structure calculations to vibrational and spectroscopic properties) and for supporting and automating advanced computational methods (such as DFT+$U$+$V$ and Wannier functions).

\section*{Results}

\subsection*{Architecture overview}

\begin{figure*}[p]
    \centering
    \includegraphics[width=\linewidth]{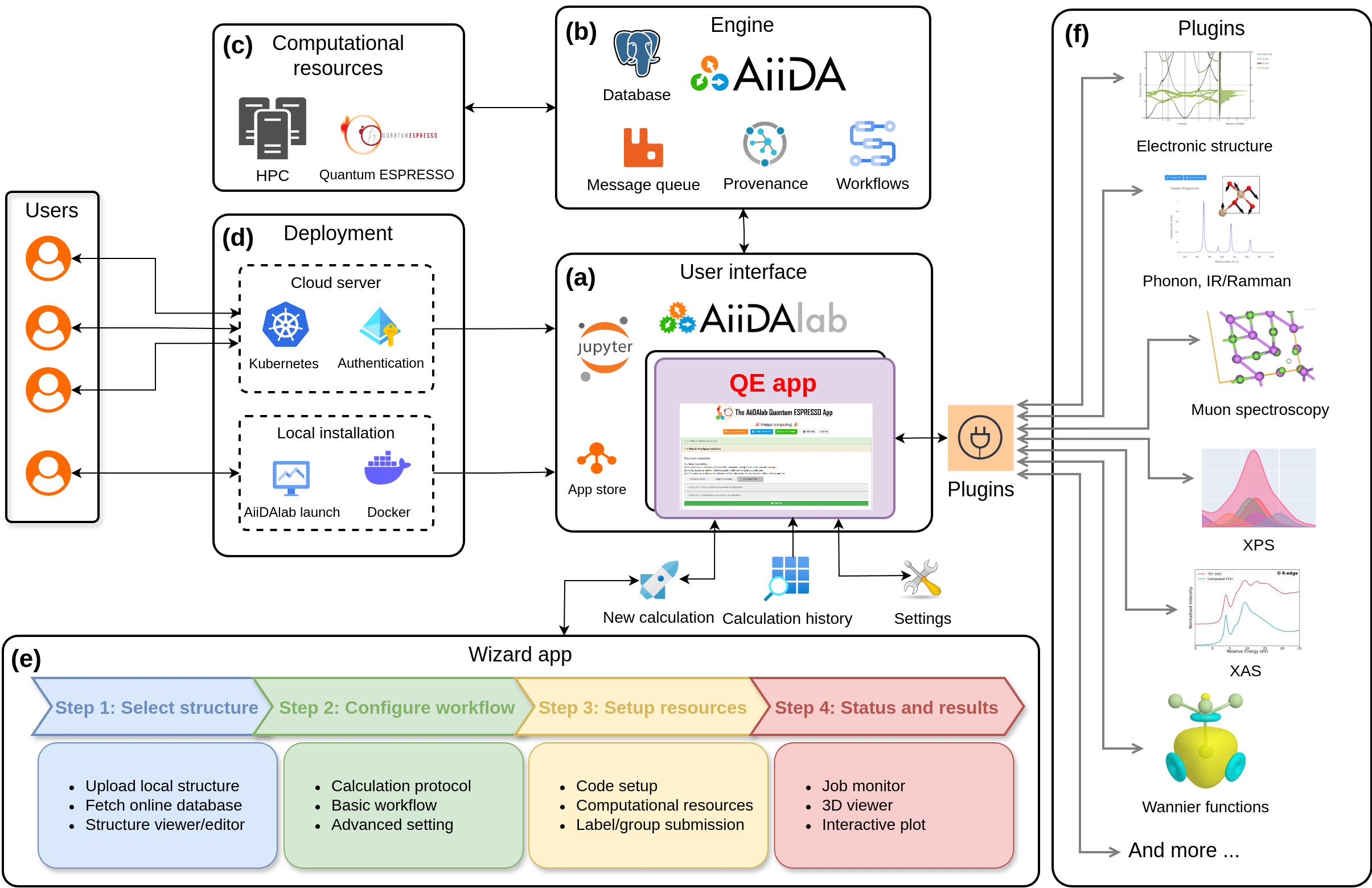}
    \caption{Overview of the QE app's architecture. 
(a) The UI is built within the AiiDAlab environment using Jupyter notebooks and the \texttt{appmode} plugin, providing an intuitive front end for simulation setup and results analysis. 
(b) The backend engine is powered by AiiDA, which handles workflow orchestration, data management, and provenance tracking. 
(c) The app connects to external high-performance computing (HPC) resources for running simulations. 
(d) The app can be deployed on cloud-based deployments using Kubernetes and authentication layers, or locally using the AiiDAlab launch tool or Docker.
(e) A modular wizard UI guides users through the four main steps of the simulation process.
(f) The plugin system enables extensibility by allowing users to activate specific property calculations or post-processing tools, such as electronic structure, vibrational properties (IR/Raman), muon spectroscopy and resting sites, XPS, XAS, Wannier functions and more. The full list of available plugins is discussed later in the text.}
    \label{fig:qeapp_architecture}
\end{figure*}

The QE app architecture aims to provide a turn-key solution for materials scientists to run DFT calculations seamlessly, addressing key challenges outlined in the introduction.
As illustrated in Fig.~\ref{fig:qeapp_architecture}, the app consists of several key components to handle different aspects of the simulation workflow: (a) a user interface (UI) built within the AiiDAlab ecosystem using Jupyter notebooks and the appmode plugin~\cite{appmode2025}, (b) an AiiDA-based backend engine responsible for workflow management and data provenance that interacts with (c) external HPC resources, (d) flexible deployment options, including both cloud-based and local installations, (e) a “wizard” app guiding users through the simulation pipeline, and (f) a plugin system for extensibility, providing access to several advanced workflows to compute different material properties, combined with specialized post-processing tools.

One of the key challenges in designing the QE app is that different material properties require distinct inputs, workflows, and outputs. A straightforward but suboptimal approach would be to create a dedicated UI or sub-app for each property, but this leads to a fragmented user experience and increased maintenance complexity. In contrast, this app adopts a unified design strategy by identifying a common interface applicable to all types of DFT calculations.
At the core of the app's UI design is the Input-Process-Output (IPO) model~\cite{goel2010computer}, which is widely used in workflow management systems, where it provides a clear and organized approach to handling tasks, ensuring systematic processing across different stages~\cite{crusoe2022methods, griem2022kadistudio}. 
The app implements the IPO model using a wizard UI that guides users through the simulation process in a structured and intuitive way (see Fig.~\ref{fig:qeapp_architecture}e). The wizard ensures that all necessary inputs are provided before proceeding to the next step, but also allows users to revisit and modify previous steps if needed, offering both flexibility and control during the setup process. The details of the wizard UI are provided in a later section.

In academic development projects, rapid prototyping often requires incorporating new algorithms and protocols into existing workflows. The QE app addresses this need with a plugin-based architecture that allows flexible addition of new functionalities. Material property calculations, such as band structures or projected density of states (PDOS), are implemented as plugins, each operating as a self-contained unit within the IPO model, managing its own input, process, and output. The main app provides a unified API and mounts the components of the plugins, ensuring seamless integration. This design preserves a consistent interface across all functionalities, allowing the main application to focus on core features shared by all plugins. It also simplifies the work of plugin developers, who can build and test their components independently, without needing to modify the core application. Plugins can be developed and maintained either within the main code base or as standalone packages. By extending functionality through plugins without altering the core structure, the QE app maintains a single, streamlined interface for setup, execution, and analysis, reducing the learning curve and delivering a smooth, consistent user experience.

\subsection*{App accessibility}
The QE app can be used both locally and via the cloud, offering flexibility for different user needs and environments. For local workstations, the app can be installed via the \texttt{aiidalab-launch} utility (see the \textbf{Code availability} section), which automates the process of retrieving the appropriate Docker image and launching a containerized instance of AiiDAlab with the QE app pre-configured. This option enables users to run the application in a fully isolated and reproducible environment on their own machines.

For cloud access, AiiDAlab's centralized Software-as-a-Service (SaaS) model allows users to access the platform via a web browser without the need for local software installations. This model supports scalable deployment, from individual users and small research groups to larger institutional infrastructures. We provide a publicly accessible demonstration instance of the QE app at (\url{https://demo.aiidalab.io}). The server is deployed via a dedicated Kubernetes infrastructure on the Microsoft Azure service and comes with the app and its core plugins pre-installed. Users can readily explore the app's functionality, inspect pre-computed examples, and submit test calculations.

\subsection*{Leveraging AiiDA for workflow management}

Workflows help standardize simulations, minimize manual intervention, and improve reproducibility, making them an essential component of the app's turn-key solutions. To manage its calculation workflows, the QE app leverages the AiiDA workflow management system, which serves as the backend engine for job execution, data provenance, and automated workflow coordination. AiiDA manages communications with HPC machines, transferring files, interacting with the job scheduler, and submitting and monitoring calculations. Besides, AiiDA supports a broader ecosystem of community-maintained plugins, available through an official registry (\url{https://aiidateam.github.io/aiida-registry/}), which covers a wide range of simulation codes, including both calculation interfaces and automated workflows.

The app uses the \texttt{aiida-quantumespresso} plugin package (see the \textbf{Code availability} section), which provides a collection of pre-configured AiiDA calculation plugins to prepare the input and parse the output of several of the \textsc{Quantum ESPRESSO} executables. These calculation classes serve as building blocks for more complex workflows that link multiple computations together, enabling users to perform sophisticated simulation tasks with minimal manual input.

AiiDA also supports automatic restarts and error handling mechanisms. Building on this, the \texttt{aiida-quantumespresso} plugin implements specific recovery strategies tailored to QE calculations. For example, if a calculation fails to reach electronic convergence, the workflow can automatically adjust key parameters, such as reducing the mixing ratio for self-consistent iterations. Similarly, if a job is interrupted due to exceeding the allocated walltime, it can be resumed cleanly using the last output structure and charge density. These recovery strategies are embedded in the workflow logic and require no user intervention, reducing manual troubleshooting and increasing the reliability of complex simulations.
In addition, AiiDA implements a caching mechanism that allows the reuse of results from previous calculations submitted with identical inputs, thus avoiding redundant computations and promoting consistency across multiple runs.
To streamline the process of preparing simulations, the app exposes other utilities from \texttt{aiida-quantumespresso} to facilitate simulation setup and customization, for example to automate $\mathbf{k}$-point mesh generation tailored to the size of the selected system.

\subsection*{QE app wizard UI}

\begin{figure}[p]
    \centering
    \includegraphics[width=\linewidth]{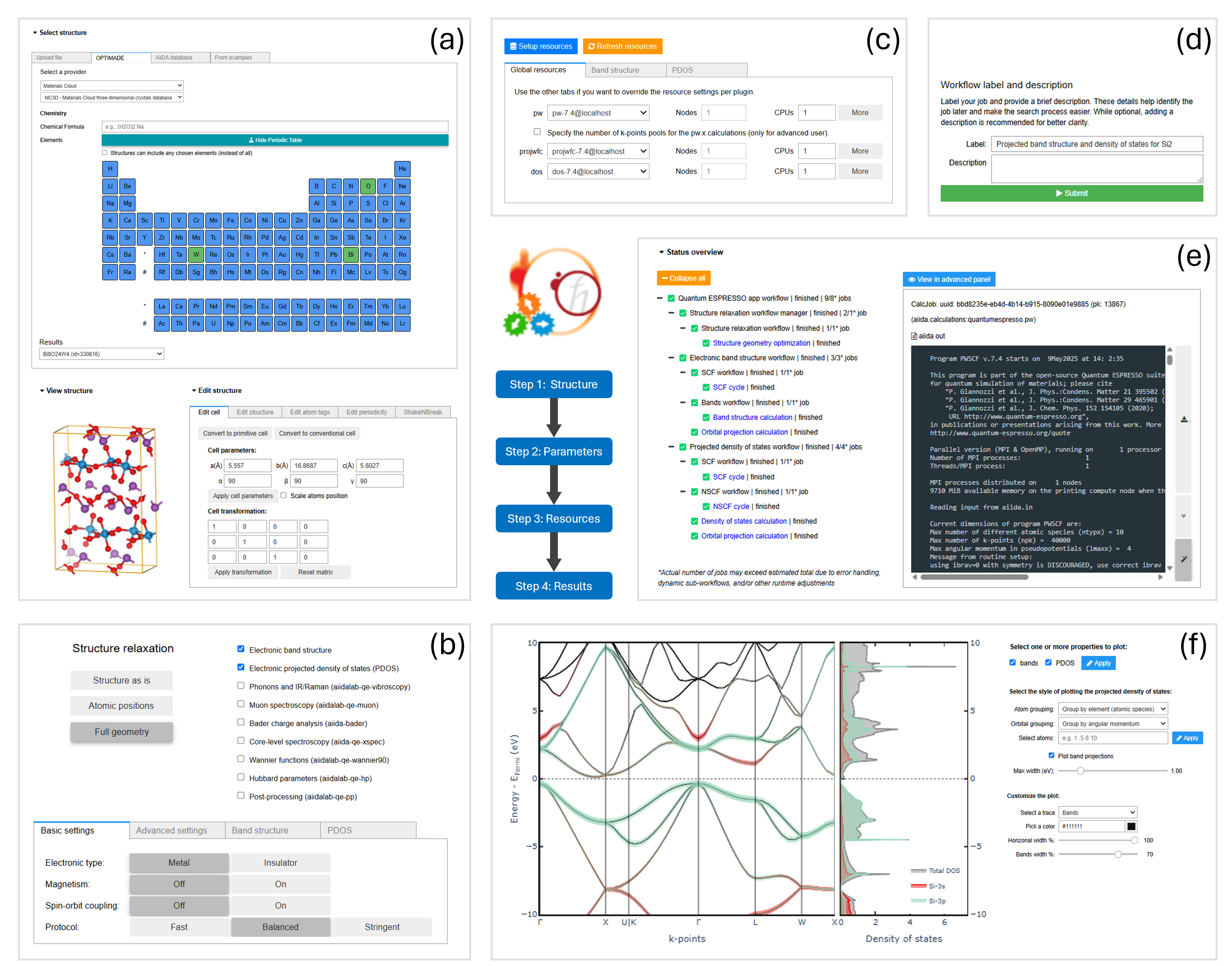}
    \caption{The QE app's wizard UI steps: (a) structure selection step, where users can select, view, and edit the input structure of the calculation; (b) workflow configuration step, where users specify the main workflow steps (structure relaxation and properties to compute) and customize the parameters of each of them; (c-d) resource selection and workflow submission step, where users select codes and computational resources, provide optional metadata, and submit the workflow; and (e-f) monitoring and analysis step, where users can monitor the running calculations, plot results from completed calculations, and customize plots interactively.}
    \label{fig:ui_workflow}
\end{figure}

The app's wizard UI serves as a guide for users as they construct and submit a workflow. Throughout the process, the wizard ensures that all necessary information is provided before progressing. Each step in the wizard consists of multiple panels that allow users to refine their input as needed. Prior to submission, users can also return to a previously confirmed step to modify any input data. The UI divides the entire process into four main steps:

\begin{enumerate}
    \item \textbf{Selecting a structure}: This step includes tools to select, view, and modify the input structure (Fig.~\ref{fig:ui_workflow}a). Users can upload their own structure (in any standard format, such as XYZ, CIF, etc.), or choose a structure from online databases compliant with the OPTIMADE standard~\cite{Andersen:2021_OPTIMADE,Evans2024} (such as the Materials Project~\cite{MaterialsProject_2013}, the Materials Cloud~\cite{talirz2020materials} and the Alexandria Materials Database~\cite{schmidt_crystal_2021, schmidt_machinelearningassisted_2023}), from their own local AiiDA database, or from a set of example structures. The structure viewer allows for inspection,  
    providing an overview of structure properties (cell parameters, volume, space/point group)
    and with a set of tools to interact with and/or manipulate the atomistic model: edit the structure cell (primitive, conventional, and supercell transformations), periodicity, (surfaces, molecule definition), and atoms (removal, tagging). Other editing features include structural modifications, such as introducing point defects in 2D materials and applying bond distortions or random displacements to explore metastable configurations of point defects, following the approach implemented in the ShakeNBreak package\cite{Mosquera-Lois2022}.

    \item \textbf{Configuring the workflow}: With an input structure selected, users can now configure the workflow, optionally including a structure geometry optimization step, selecting from a host of properties of interest to compute (see \textbf{Plugins} section, also in Supplementary Information), and customizing calculation parameters (Fig.~\ref{fig:ui_workflow}b). To simplify this process, the app splits the parameter customization step into basic, advanced, and property-specific panels. The basic panel streamlines the process by abstracting much of the complexity via a set of top-level controls, including a protocol selector to set most input parameters balancing calculation speed and accuracy~\cite{nascimento2025}. Full flexibility in customizing calculation inputs is retained via the advanced panel. In this panel, the user can choose among various pseudopotential families, including the Standard Solid-State Pseudopotentials (SSSP; efficiency or precision)~\cite{prandini_precision_2018} and PseudoDojo (standard or stringent)~\cite{van_setten_pseudodojo_2018}, or alternatively upload custom pseudopotential files directly via the GUI. Further details of each panel are provided in the \textbf{Workflow parameter settings panels} section of the Supplementary Information.

    \item \textbf{Choosing computational resources and submitting the workflow}: Next, users proceed to choose the computational resources to use by AiiDA to run the calculation (Fig.~\ref{fig:ui_workflow}c). The app provides a straightforward UI for selecting AiiDA \texttt{Code} instances---references to executables on local or remote machines---and specifying the computational resources to be allocated per code (e.g., nodes, CPUs). Additional codes may be setup in a separate notebook (see the \textbf{Computational resources setup} section of the Supplementary Information). Once the resources have been chosen, users may customize the auto-generated workflow label and optionally provide a description to attach to the workflow (Fig.~\ref{fig:ui_workflow}d). When ready, users can submit the workflow. AiiDA will handle submission in the background.

    \item \textbf{Monitoring and viewing workflow results}: After submitting a workflow, the app redirects users to the status panel of the results step (Fig.~\ref{fig:ui_workflow}e), where they can monitor the calculations in real time. Users can switch to the summary panel to inspect a summary of the selected inputs. As calculations complete, results from each calculation submitted by the app's workflow are made available for analysis in dedicated sub-panels of the results panel (Fig.~\ref{fig:ui_workflow}f). Lastly, when the workflow is finished, users can download from the summary panel either a zip file including the raw input and output data of all workflow steps, or directly a single AiiDA archive (including the full provenance graph) for further inspection or sharing. The latter options (AiiDA archive) allows users to import the workflow into another AiiDAlab instance and continue inspecting the results via the QE app GUI.
\end{enumerate}

The app's plugin-based architecture provides several entry points to extend the core UI described above. Plugin developers can introduce additional UI components, including structure importers and editors, property-specific parameter settings panels, computational resource panels to override global resources per property-specific code, and property-specific results panels. The following section provides examples of such plugin components.

\subsection*{Plugins}

The QE app's plugin-based architecture provides a flexible framework for extending the platform with specialized functionalities, ranging from electronic-structure analysis to advanced spectroscopic techniques. Each plugin incorporates its own workflows and UI components, yet remains fully integrated within the app's unified IPO model and the broader AiiDAlab environment. This modular design not only streamlines maintenance and testing but also encourages community-driven extensions. Below, we highlight the key aspects of two plugins, one provided as a core plugin included in the app's code base (electronic structure) and one developed as an external plugin (vibrational spectroscopies). We also provide a short summary for each of the currently supported app plugins. The plugin packages are listed in the \textbf{Code availability} section.

\subsubsection*{Electronic structure: band structures and (projected) densities of states}
Electronic band structure theory is fundamental for interpreting materials behavior including electrical transport and optical properties, making it a cornerstone of solid-state physics~\cite{GROSSO2014179}. In the QE app, band-structure calculations are streamlined through a dedicated plugin that guides the users from processing to final visualization. The process begins with supplying an initial crystal structure, which can be either a representation of an experimentally known structure, or a computationally optimized one. The app automatically defines the band-structure path based on the structure. For 3D materials, the app leverages existing tools---such as seekpath~\cite{HINUMA2017}---to standardize the lattice and identify a $\mathbf{k}$-point path based on well-known symmetry lines~\cite{Togo31122024}. In low-dimensional cases (e.g., 2D monolayers), the app detects the system symmetry and selects the appropriate reciprocal paths that capture the most relevant dispersion features.

\begin{figure}[h]
    \centering
    \includegraphics[width=8cm]{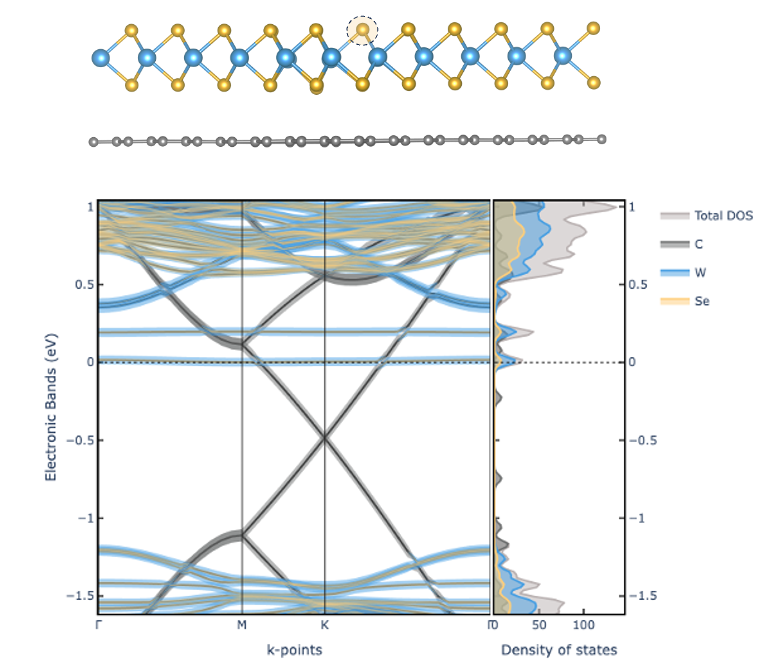}
    \caption{Schematic illustration of a selenium (Se) vacancy in a \ch{WSe2}/graphene heterojunction (top) and the corresponding projected band structure and projected density of states (PDOS) (bottom), computed using the QE app with spin-polarized DFT using the PBE exchange-correlation functional\cite{PBE}, and including SOC. This study explores the hybridization of Se vacancy states with graphene and their interactions. The in-gap defect states, which are non-dispersive and exhibit a gap due to spin–orbit coupling (SOC) effects, are located between the conduction and valence bands of \ch{WSe2}, as discussed in Ref.~\citenum{bobzien2024}.
   }
    \label{fig:band_pdos_plugin_example}
    \end{figure}

Beyond computing the band energies, the plugin incorporates projected band structure analysis, also known as “fat bands”, which quantifies the contribution of specific atomic orbitals to each energy level; an example is shown in Fig.~\ref{fig:band_pdos_plugin_example}f. These projections help reveal crucial effects such as orbital hybridization, localized defect states, or spin-splitting phenomena\cite{schuler2019}.
In addition to fat bands, where projection onto atomic orbitals is studied only along the high-symmetry paths in reciprocal space where the band structure is computed, projected density of states (PDOS) calculations provide an additional perspective on the electronic structure of materials by revealing how specific atoms or orbitals contribute to the total density of states (DOS), e.g. helping in identifying which atomic orbitals dominate states near the Fermi level, thereby offering insights into the origins of electronic, magnetic, or optical properties.
In the QE app, the PDOS workflow integrates four core steps: self-consistent field (SCF), non-self-consistent field (NSCF), total DOS, and PDOS, into a streamlined sequence that automatically generates necessary wavefunction files and appropriately handles large intermediate data. 
In particular, orbital-projected DOS calculations decompose the total DOS into site- and orbital-specific contributions, highlighting the role of, for instance, $d$-orbitals in transition-metal oxides or $p$-orbitals in organic semiconductors. 
Moreover, both the band structure and PDOS plugins support spin-orbit coupling (SOC) calculations, enabling an accurate description of relativistic effects in materials where SOC plays a significant role, such as heavy-element semiconductors and topological insulators, and allowing users to analyze SOC-induced modifications in the DOS and PDOS.
Within the standard app UI, users can also configure advanced parameters, such as $\mathbf{k}$-point densities, smearing schemes, or energy ranges. This flexibility is particularly important for materials where fine spectral features can strongly influence theoretical predictions of conductivity or band alignments.
The final PDOS data are presented via interactive plots, allowing for immediate visual inspection of peaks and energy offsets relative to the Fermi level. The PDOS can also be displayed in tandem with the band structure, further clarifying which states govern conduction or valence behavior, as we illustrate in Fig.~\ref{fig:band_pdos_plugin_example}.
The interactive panel of the plugin enables users to adjust visual elements, isolate specific states, and export publication-quality figures or the underlying numerical data. By automating complex workflows in a reproducible environment, this plugin significantly reduces the computational barrier for advanced electronic structure studies of complex materials.

\subsubsection*{Vibrational spectroscopies: phonons and Raman/Infrared spectra}
Vibrational spectroscopies are among the most powerful methods for materials structure investigation, via fingerprint atomic vibrations probed by using several types of radiation, such as light.
The simulation of vibrational properties in the QE app is available through the \texttt{aiidalab-qe-vibroscopy} plugin. 
The app has been successfully applied to investigate the vibrational and optoelectronic properties of \ch{BaZrS3}, highlighting its practical utility in ongoing research and materials characterization efforts~\cite{nielsen2025}.
This is accomplished via the specific workflow, called \texttt{VibroWorkChain}, which orchestrates the submission of AiiDA workflows provided by the \texttt{aiida-vibroscopy}~\cite{bastonero_automated_2024}
and \texttt{aiida-phonopy}~\cite{bastonero_automated_2024, aiida_phonopy_github, phonopy-phono3py-JPCM, phonopy-phono3py-JPSJ} plugins. 
Specifically, phonon properties are obtained using a finite-difference approach~\cite{Togo31122024} through multiple SCF calculations,
involving finite atomic displacements from equilibrium positions in supercells. Large enough supercells are needed to accurately interpolate phonon dispersions, equivalently to using dense \textbf{q}-point meshes in density-functional perturbation theory (DFPT)~\cite{Baroni-DFPT-2001}. If long-range non-analytic corrections (NACs) are required, the workflow will also compute dielectric properties via finite applied electric fields~\cite{Umari2002} simulations. This approach differs from the one implemented in the \texttt{PHonon} code of the \textsc{Quantum ESPRESSO} package~\cite{Baroni-DFPT-2001}, and enables calculations with any functional, such as hybrid functionals or DFT+$U$(+$V$)~\cite{bastonero_automated_2024}.
Various types of simulations can be performed based on the desired outcome: (1) full characterization of phonon properties, including phonon bands, phonon PDOS, thermodynamics, and infrared (IR) and Raman spectra; (2) phonon properties for non-polar materials (i.e., NACs are not computed); (3) IR/Raman spectra from $\Gamma$-point normal modes of the unit cell;
(4) dielectric properties (dielectric and Raman tensors, Born effective charges).
As shown in Supplementary Fig. S3, options are provided to users by means of a drop-down menu, together with options to select (or automatically determine) the supercell size needed for finite atomic displacements and the symmetry tolerance used to detect unique atomic displacements (thus reducing the cost of the simulation). An estimation of the number of finite displacement supercells can be performed, providing an indication of the number of simulations that are going to be submitted, and thus of the total computational cost.

If phonons are computed (i.e., if users select one of the first two calculation options), the inelastic neutron scattering (INS) dynamical structure factor can additionally be computed as a live post-processing functionality in the output panel. 
Specifically, INS observables for both single-crystal and powder samples are calculated by means of the \texttt{Euphonic} package~\cite{fair_euphonic_2022} in a format directly 
comparable with experimental spectra. Users also have the possibility to provide a specific reciprocal space path or plane to facilitate comparison of the simulation results with specific experimental setup and measurements.

\begin{figure}[h]
    \centering
    \includegraphics[width=\linewidth]{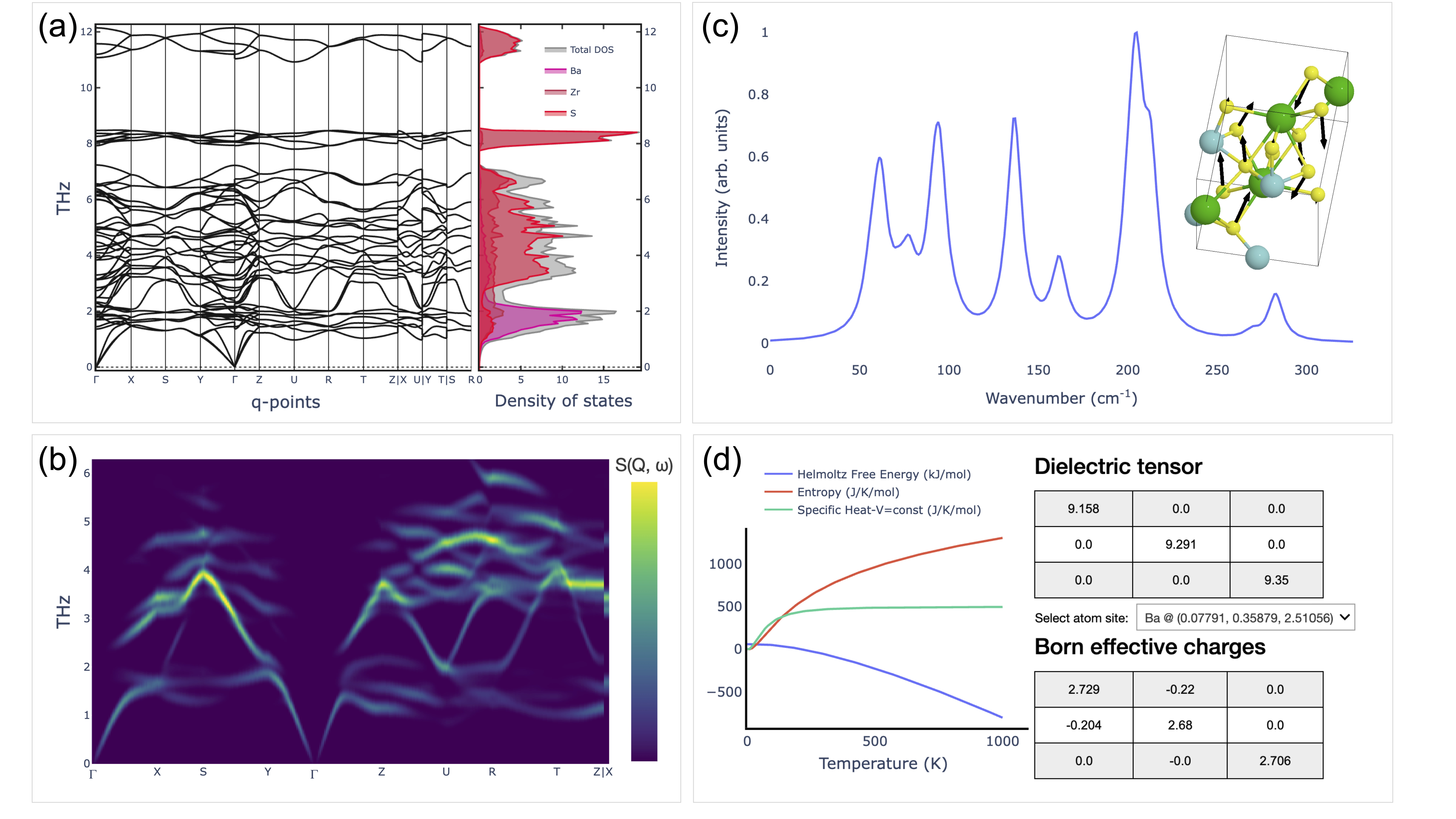}
    \caption{Results panels for the aiidalab-qe-vibroscopy plugin. (a) Phonon dispersion plot for \ch{BaZrS3}, together with the phonon DOS (aligned in energy with respect to the phonon bands). (b) Computed INS spectrum. (c) Single crystal Raman spectrum and modes animation. (d) Thermodynamic properties, dielectric tensor and Born effective charges.}    \label{fig:vibro_example}
    \end{figure}

Selected screenshots of the plugin interface are shown in Fig.~\ref{fig:vibro_example} for the previously mentioned \ch{BaZrS3} system~\cite{nielsen2025}, showing that the results tab is subdivided into different sub-tabs, each of them dedicated to a specific output of the simulations:
(i) phonon band structure, phonon  PDOS and thermodynamic quantities, see panel (a); (ii) INS results, to be computed live, see panel (b); (iii) IR and Raman active modes, together with their interactive 3D animations, see panel (c); (iv) thermodynamic and dielectric properties, see panel (d). All the data shown in the panels can be downloaded and analyzed with additional external tools, if needed. 

\subsubsection*{Additional plugins}

\begin{figure}
    \centering
    \includegraphics[width=\linewidth]{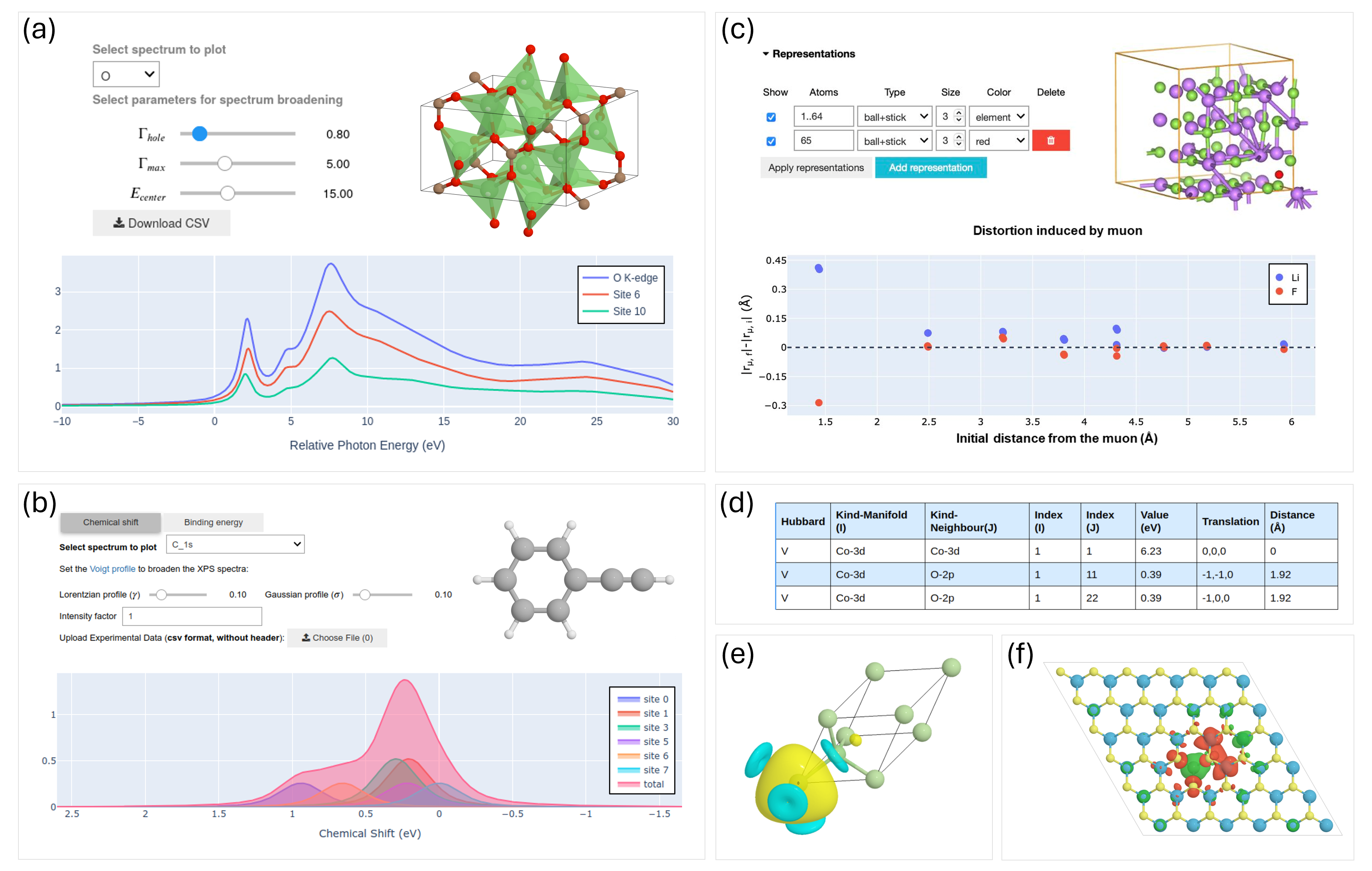}
    \caption{Overview of result panels from various plugins. (a) XAS plugin showing the O K-edge XANES spectrum of crystalline \ch{Li2CO3}, including site-resolved and total contributions. (b) XPS plugin applied to phenylacetylene (\ch{C8H6}) in gas phase, illustrating the \ch{C} 1s core level shifts and their relation to atomic environments. (c) Muon spectroscopy plugin showing a candidate muon site in \ch{LiF} and the distortions induced by the muon to the hosting lattice. (d) Hubbard $U\!+\!V$ plugin calculating on-site and inter-site interactions in \ch{LiCoO2}, including a table of computed parameters and their atomic associations. (e) The Wannier function plugin  showing isosurfaces of one of the maximally localized Wannier functions in gallium arsenide. (f) Post-processing plugin visualizing spin density data in a sulfur vacancy in a monolayer of \ch{MoS2}.}
    \label{fig:plugins}
\end{figure}

Beyond the two headline examples, the app supports a growing suite of specialized plugins designed to automate and visualize advanced materials simulations. Below, we summarize the key functionalities and materials properties addressed by each plugin. Representative result panels from these plugins are shown in Fig.~\ref{fig:plugins}, while setup options and further technical details are provided in the \textbf{Plugins} section of the Supplementary Information.

\begin{itemize}
    \item \textbf{X-ray absorption spectroscopy:} the XAS plugin facilitates the computation of X-ray absorption near-edge structure (XANES), a technique sensitive to the local chemical environment of absorbing atoms~\cite{Bokhoven2016}. The plugin is based on a reciprocal-space pseudopotential scheme, as implemented in the XSpectra code~\cite{Gougoussis2009,Bunau2013}, and accounts for different core-hole treatments, symmetry-analysis for inequivalent site selection, and lifetime broadening~\cite{Bunau2013}. Users can select the absorbing element and inspect both total and site-resolved XANES spectra. Interactive controls allow adjusting broadening parameters and exporting spectra for further analysis.
    
    \item \textbf{X-ray photoelectron spectroscopy:} this plugin enables the calculation of core-level binding energies using the Delta Kohn-Sham ($\Delta$KS) approach~\cite{triguero1998calculations, CAVIGLIASSO1999,walter2016offset}, a method widely used to interpret experimental XPS spectra and identify chemical shifts across atomic sites. Users can select target atoms, specify core-hole pseudopotentials, and set spectral broadening parameters. The results panel displays both the computed binding energies and their site-resolved chemical shifts in an interactive table and spectrum viewer, allowing comparison with experimental reference data.

    \item \textbf{Muon spectroscopy:} this plugin targets the prediction of muon stopping sites and their interaction with magnetic environment, crucial for interpreting muon spin relaxation ($\mu$SR) experiments ~\cite{Dalmas+gen+1997}. By simulating a dilute positive muon in a material using DFT+$\mu$~\cite{Moller+fluoridessites+2013,bernardini+muonsites+2013,Blundell2025}, users can identify likely stopping sites and compute associated local fields. The setup panel lets users configure the supercell size, charge state, and number of trial sites (inputs for the \texttt{aiida-muon} plugin~\cite{aiida-muon-plugin,aiida-muon-zenodo}). The results include a ranked list of candidate sites with respect to relative total energy, 3D site visualization and, optionally, the muon spin polarization functions computed for different crystal orientations via the \texttt{UNDI} package~\cite{BONFA2021107719}.

    \item \textbf{Wannier functions:} the Wannierization plugin automates the generation of maximally localized Wannier functions (MLWFs), used to analyze bonding characters, build tight-binding models, and study topological invariants~\cite{marzari1997maximally,Marzari2012,MarrazzoRMP2024}. Users can choose between automated Wannierization schemes [e.g., the selected columns of the density matrix (SCDM) method~\cite{damle2015compressed,Vitale2020} or the projectability disentanglement Wannier functions (PDWF) method~\cite{qiao2023projectability}] and adjust the energy windows\cite{qiao2023projectability}. It is also possible to compute the Fermi surface using Wannier interpolation, allowing for dense Brillouin zone sampling (k-point distance down to a few hundredths of reciprocal angstrom). For the generated Fermi surface, it is possible to calculate de Haas--van Alphen (dHvA) oscillation frequencies, facilitating direct comparison with experimental measurements  \cite{Biberacher2005, Bergemann2005}. The result panel compares DFT and Wannier-interpolated bands, reports total and component spreads, offers 3D visualization of individual Wannier functions with atoms, and provides a plot with computed dHvA frequencies. Files containing the tight-binding model in the MLWF basis set and the Fermi surface are available in the “Downloads” section of the result panel.

    \item \textbf{Hubbard parameters:} accurate determination of on-site $U$ and inter-site $V$ Hubbard corrections is essential for treating transition-metal and rare-earth compounds containing partially filled and localized $d$ and/or $f$ electrons~\cite{anisimov:1991, Liechtenstein:1995, Dudarev:1998, campo2010extended}. This plugin interfaces with the \texttt{aiida-hubbard} workflows~\cite{bastonero2025hubbard}, which efficiently orchestrate DFPT-based calculations~\cite{Timrov:2018, Timrov:2021} to compute these parameters from first principles using the \texttt{hp.x} code of QE\cite{timrov2022hp}. Users can toggle between ``one-shot'' or ``self-consistent'' modes, choose atom pairs for inter-site corrections, and specify convergence thresholds. Results include a summary table of the computed Hubbard parameters and an interactive structure viewer linking each value to its corresponding site or atom pair.

    \item \textbf{Post-processing:} the post-processing plugin enables real-space visualization of properties such as charge and spin densities, wavefunctions, electrostatic potential, integrated local density of states (ILDOS), local density of states at specific energies, and STM images using the \texttt{pp.x} and Critic2 codes~\cite{otero_critic2014}. Users can select previously completed calculations, choose the target quantity, and define isosurface or plane-plot parameters. Results are displayed using an integrated 3D viewer powered by \texttt{weas-widget}~\cite{weas_widget_github}, providing an intuitive way to correlate electronic structure features with spatial distributions.

    \item \textbf{Bader charge analysis:} by partitioning the charge density into atomic basins, the Bader plugin (based on the Bader code~\cite{tang2009grid}) provides chemically meaningful net atomic charges---useful for understanding charge transfer, oxidation states, or catalysis. The plugin processes charge density data from completed calculations and presents the atomic charges in a sortable, interactive table alongside the structural model.

\end{itemize}

\subsection*{App utilities}

Apart from the core and plugin functionalities encapsulated in the wizard UI, the QE app also provides a set of utility pages for the following operations:

\begin{itemize}
    \item \textbf{Browsing calculation history}: Users can browse, filter, and manage all previously submitted jobs (see Supplementary Fig. S14). 
    
    \item \textbf{Installing app plugins}: Users can find, install, and manage plugins, customizing their computational environment as needed (see Supplementary Fig. S15).
    
    \item \textbf{Configuring computational resources}: Users can define local and remote machines and codes, as well as browse (and optionally disable) existing codes (see Supplementary Fig. S16).
    
    \item \textbf{Downloading example calculations}: Users can import example AiiDA archives into their database with one click. We provide a host of examples spanning core and plugin functionality, covering most common use cases (see Supplementary Fig. S17).
\end{itemize}

Further details on each utility are provided in the \textbf{Utilities} section of the Supplementary Information.

\section*{Discussion}

The QE app and its underlying AiiDA ecosystem demonstrate in practice how the FAIR principles can be extended beyond data only, to include the full life-cycle of a first-principles simulation, including the workflows that generate the data, the codes that execute them, and the interactive tools used to analyse the results~\cite{blum2024roadmap}. Below, we map each FAIR dimension onto concrete features that are already available in the QE app or planned for the near future, underscoring the added value for both end-users and developers.

\textbf{Findable}: Every calculation launched from the QE app is automatically stored in the AiiDA provenance graph. Users can export the complete graph---including inputs, outputs and metadata---as a single AiiDA archive file. These archives can be deposited in the \emph{Materials Cloud Archive}~\cite{talirz2020materials}, which offers persistent DOIs and a rich web interface that lets third parties browse the full data lineage and inspect input and output files. At the workflow layer, AiiDA's entry-point mechanism registers every workflow, parser and calculation plugin with a unique name and provides a plugin registry for plugin developers to register their workflows. At the plugin layer, the app provides a \emph{Plugin management} page, where users can discover and install additional plugins from the community.

\textbf{Accessible}: The QE app is available in both cloud-based and local configurations, lowering the entry barrier for non-specialists. In the cloud, it can be accessed directly through a web browser without installation or setup. The cloud deployment is managed by Kubernetes and JupyterHub, which handle authentication and authorization via standard OAuth2 protocols, ensuring secure and scalable access. To provide an immediate hands-on experience, we also offer a publicly accessible demonstration instance at \url{https://demo.aiidalab.io}. For local use, the app can be launched using the \texttt{aiidalab-launch} utility, which sets up a pre-configured, containerized environment. These deployment options (see App accessibility section) ensure flexibility and broad accessibility across user environments.

\textbf{Interoperable}: While the app currently supports only \textsc{Quantum ESPRESSO}, its architecture is designed to be interoperable by serving as a generic template that can be adapted to other DFT codes. By exploiting the common workflow interfaces for over ten DFT codes that have been defined and implemented in AiiDA~\cite{huber2021common,bosoni2024verify}, developers can reuse the app's structure to build similar applications for other quantum engines. For example, we are currently exploring adapting this framework to the CP2K code by integrating the independently developed \texttt{aiidalab-empa-surfaces} app~\cite{aiidalab_empa_surfaces} into the same structural architecture used in the app. The app's architecture is also serving as a model for a DFT-code-agnostic GUI developed as part of the PREMISE project of the ETH Board (https://ord-premise.org/). Furthermore, the interactive widgets and results analysis tools integrated into the app have proven invaluable not only for non-specialists but also for seasoned computational materials scientists. To maximize their utility, we plan to further modularize these tools, allowing users to employ them outside of the app's main interface, such as in a standard Jupyter notebook. These tools can also be utilized on other platforms, like the Materials Cloud website~\cite{talirz2020materials} and the OSSCAR platform~\cite{du2023osscar,Du2024}.

\textbf{Reusable (and Reproducible)}: All components of the QE app, including the main interface, the underlying AiiDA plugins, and the analysis widgets, are released under permissive open-source licenses and hosted on public GitHub repositories (see the \textbf{Code Availability} section), maximizing reusability and transparency.
By using AiiDA for workflow management, the app ensures that all inputs, processes, and outputs are automatically stored in a structured,  queryable database. When a calculation is restarted (e.g., to resolve SCF convergence issues), the updated settings are also recorded, preserving the full decision-making process. Even publication-quality figures generated via the interactive plotting widgets are fully reproducible with a single click, as the complete widget state (e.g., axis ranges, color maps, structure viewer settings) is stored alongside the corresponding AiiDA node. While reproducibility goes beyond the original FAIR acronym, our design makes it a practical reality and aligns with the provenance-focused spirit of the FAIR guidelines. Furthermore, the app's plugin architecture ensures extensibility and modularity, allowing users and developers to share and reuse specialized workflows for diverse material properties and advanced analytical techniques, thus facilitating broader community collaboration.

Apart from FAIR, the development of the QE app also underscores the importance of user-friendly interfaces. These interfaces lower entry barriers for non-specialists and accelerate the onboarding process. For example, the in-app guide system, which overlays context-specific information and “click-here-next” instructions on top of any panel, trains the user in place, avoiding the need for external manuals (see Supplementary Fig. S13). Ongoing development efforts aimed at decoupling the UI from the AiiDA backend via modern UI frameworks and dedicated REST APIs will significantly enhance scalability, multi-user capabilities, and collaborative potential. Such architectural evolution will enable more efficient use of computational resources, broader access, and robust support for collaborative research initiatives, further solidifying the app as a cornerstone platform in the FAIR simulation landscape.

The QE app represents a significant advancement in making DFT simulations more accessible to the broader communities by offering turn-key solutions that simplify the complexities of computational materials science. With its intuitive GUI, the app guides users through the simulation workflow---from structure selection to results analysis---reducing barriers for non-specialists. Its modular plugin-based architecture supports diverse materials-science workflows and allows for rapid integration of new functionalities, keeping the app up to date with ongoing developments in computational materials science. In conclusion, the QE app lowers the barrier to advanced DFT simulations, enabling a broader range of researchers to adopt computational tools and workflows, accelerating collaboration and driving forward materials research.

\section*{Methods}

\subsection*{Technical details}

Given the complexity of the QE app's UI and underlying logic, careful architectural choices are essential to ensure long-term maintainability and extensibility. To this end, the app adopts the Model-View-Controller~\cite{krasner1988mvc, gamma1994design} (MVC) design pattern, which provides a clear separation between the UI, the underlying data models, and the computational logic. This structure enhances scalability, simplifies testing, and improves responsiveness by enabling lazy-loading techniques. The details of the MVC implementation, including the use of the Observer and Mediator supporting design patterns via the \texttt{traitlets} package, and the model network architecture, are provided in the \textbf{Model-View-Controller} section of the Supplementary Information.

\subsection*{Computational details}
All DFT calculations in this work are performed using the default settings of the \texttt{aiida-quantumespresso} balanced protocols, as defined in Ref.~\citenum{nascimento2025}. The SSSP v1.3.0 efficiency pseudopotential family~\cite{prandini_precision_2018, prandini_2023_rcyfm-68h65} is used throughout. For the phonon simulations of \ch{BaZrS3} shown in Fig.~\ref{fig:vibro_example}, a $2 \times 2 \times 2$ supercell is employed to compute the force constants, which are necessary to obtain the phonon dispersion and related vibrational properties.

\section*{Code Availability}

The QE app is available as an open-source project and can be accessed from its GitHub repository: \url{https://github.com/aiidalab/aiidalab-qe}. The \texttt{aiida-quantumespresso} plugin and its associated workflows are hosted at \url{https://github.com/aiidateam/aiida-quantumespresso}. The AiiDAlab platform, which serves as the foundation for these applications, is maintained under the AiiDAlab GitHub organization: \url{https://github.com/aiidalab}. The \texttt{aiidalab-launch} tool is hosted at \url{https://github.com/aiidalab/aiidalab-launch}.

Additionally, QE app plugins for different functionalities are available:
\begin{itemize}
    \item \textbf{aiidalab-qe-vibroscopy}: \url{https://github.com/aiidalab/aiidalab-qe-vibroscopy}
    \item \textbf{aiidalab-qe-muon}: \url{https://github.com/aiidalab/aiidalab-qe-muon}
    \item \textbf{aiidalab-qe-hp}: \url{https://github.com/aiidalab/aiidalab-qe-hp}
    \item \textbf{aiidalab-qe-wannier90}: \url{https://github.com/aiidalab/aiidalab-qe-wannier90}
    \item \textbf{aiida-qe-xspec}: \url{https://github.com/aiidaplugins/aiida-qe-xspec}
    \item \textbf{aiidalab-qe-pp}: \url{https://github.com/AndresOrtegaGuerrero/aiidalab-qe-pp}
    \item \textbf{aiida-bader}:
    \url{https://github.com/superstar54/aiida-bader}
\end{itemize}

All software mentioned in this paper is open-source and freely available to the community.

\bibliography{main}

\section*{Acknowledgments}
We gratefully thank Carl Simon Adorf for initial contributions and developments to the QE app. We thank the AiiDA team for their continuous support and contributions to the AiiDA framework, which underpins the QE app, the \textsc{Quantum ESPRESSO} developers for their ongoing efforts in maintaining and improving the QE code, which is the computational engine behind the app, and all users who tested the app and provided valuable feedback, including Nicola Colonna, Thomas J. Hicken, Jonas A. Krieger, Stanislav Nikitin and Tom Fennell.
XW, EB, MBo, AOG, MBe, DD, SPH, NP, JQ, TR, CJS, IT, AVY, JY, NM, CAP and GP acknowledge financial support by the NCCR MARVEL, a National Centre of Competence in Research, funded by the Swiss National Science Foundation (grant number 205602). 
EB, CAP and GP acknowledge financial support by the Open Research Data Program of the ETH Board (project ``PREMISE'': Open and Reproducible Materials Science Research).
EM and DP acknowledge support by the MaX European Centre of Excellence – Materials design at the eXascale (www.max-centre.eu), of which Quantum ESPRESSO is a lighthouse code; MaX is supported by the European High Performance Computing Joint Undertaking and participating countries (grant No. 101093374).
PNOG and DP acknowledge financial support by the European Union - NextGenerationEU through the Italian Ministry of University and Research under PNRR - M4C2I1.4 ICSC - Centro Nazionale di Ricerca in High Performance Computing, Big Data and Quantum Computing (Grant No. CN00000013) through the Innovation Grant ASGARD.
LB and NM acknowledge financial support by the Deutsche Forschungsgemeinschaft (DFG) under Germany's Excellence Strategy (EXC 2077, No. 390741603, University Allowance, University of Bremen) and Lucio Colombi Ciacchi, the host of the ``U Bremen Excellence Chair Program''.
NP, JY and GP acknowledge financial support by the Swiss National Science Foundation (SNSF) Project Funding (grant 200021E\_206190 ``FISH4DIET'').
IT acknowledges financial support by the Swiss National Science Foundation (SNSF) Project Funding (grant 200021-227641 and 200021-236507).
MBe, XW, JY and GP acknowledge financial support by the SwissTwins project, funded by the Swiss State Secretariat for Education, Research and Innovation (SERI).
DD and GP acknowledge financial support by the EPFL
Open Science Fund via the OSSCAR project.
XW, PNOG, MAHB, EM, DP, NM and GP acknowledge financial support by the European Union's Horizon 2020 research and innovation programme under grant agreement No. 957189 (BIG-MAP), also part of the BATTERY 2030+ initiative under grant agreement No. 957213.
JY, NM and GP acknowledge financial support by the MARKETPLACE project funded by Horizon 2020 under the H2020-NMBP-25-2017 call (Grant No. 760173).
DH acknowledges financial support by the European Union's Horizon 2020 research and innovation programme under grant agreement No. 803718 (SINDAM) and the UK Research and Innovation (UKRI) EPSRC grant ref. EP/X026973/1.
PB, RDR and IJO acknowledge financial support by the PNRR MUR project ECS-00000033-ECOSISTER and from University of Parma through the action ``Bando di Ateneo 2023 per la ricerca''. 
We acknowledge access to Piz Daint and Alps at the Swiss National Supercomputing Centre (CSCS), Switzerland under the MARVEL's share with the project IDs mr32 and mr33.
This work was further supported by grants from the Swiss National Supercomputing Centre (CSCS) under project IDs lp18, s1267 and s1295 on Alps. 
We acknowledge ISCRA, ECOSISTER and ICSC for awarding access to the LEONARDO supercomputer, hosted by CINECA (Italy).

\section*{Author contributions statement}
Conceptualization: XW, EB, MBo, AOG, AVY, JY, NM, CAP, GP;
Methodology: XW, EB, MBo, AOG, LB, MBe, PB, RDR, DD, PNOG, MAHB, DH, SPH, EM, IJO, NP, DP, JQ, TR, CJS, IT, AVY, JY, NM, CAP, GP;
Software: XW, EB, MBo, AOG, LB, MBe, PB, DD, PNOG, MAHB, DH, SPH, IJO, NP, DP, JQ, TR, CJS, IT, AVY, JY, CAP, GP;
Validation: XW, EB, MBo, AOG, DP, JY, CAP, GP;
Investigation: XW, EB, MBo, AOG, JY;
Resources: RDR, EM, DP, NM, CAP, GP;
Writing – original draft: XW, EB, MBo, AOG;
Writing – review \& editing: XW, EB, MBo, AOG, LB, MBe, PB, RDR, PNOG, MAHB, DH, EM, IJO, NP, DP, JQ, TR, IT, AVY, JY, NM, CAP, GP;
Visualization: XW, EB, MBo, AOG, PNOG, MAHB, TR, AVY, JY;
Supervision: RDR, EM, DP, NM, CAP, GP;
Project administration: NM, CAP, GP;
Funding acquisition: RDR, EM, DP, NM, CAP, GP.

\section*{Competing interests}
The authors declare no competing financial interest.

\clearpage
\begingroup
\pagestyle{empty}


\section*{Supplementary material}

# Supporting Information: Making atomistic materials calculations accessible with AiiDAlab

**Xing Wang**[1,2,†], **Edan Bainglass**[1,2,†], **Miki Bonacci**[1,2,†], **Andres Ortega-Guerrero**[3,†], **Lorenzo Bastonero**[4], **Marnik Bercx**[1,2], **Pietro Bonfà**[6,7], **Roberto De Renzi**[8], **Dou Du**[5], **Peter N. O. Gillespie**[7], **Michael A. Hernández-Bertrán**[6,7], **Daniel Hollas**[9], **Sebastiaan P. Huber**[5], **Elisa Molinari**[6,7], **Ifeanyi J. Onuorah**[8], **Nataliya Paulish**[1,2], **Deborah Prezzi**[7], **Junfeng Qiao**[5], **Timo Reents**[1,2], **Christopher J. Sewell**[5], **Iurii Timrov**[1,2], **Aliaksandr V. Yakutovich**[3], **Jusong Yu**[1,2], **Nicola Marzari**[1,2,4,5], **Carlo A. Pignedoli**[3,*], and **Giovanni Pizzi**[1,2,3,*]

[1]PSI Center for Scientific Computing, Theory and Data, 5232 Villigen PSI, Switzerland
[2]National Centre for Computational Design and Discovery of Novel Materials (MARVEL), 5232 Villigen PSI, Switzerland
[3]nanotech@surfaces Laboratory, Empa-Swiss Federal Laboratories for Materials Science and Technology, 8600 Dübendorf, Switzerland
[4]U Bremen Excellence Chair, Bremen Centre for Computational Materials Science, and MAPEX Center for Materials and Processes, University of Bremen, 28359 Bremen, Germany
[5]Theory and Simulation of Materials (THEOS), École Polytechnique Fédérale de Lausanne, 1015 Lausanne, Switzerland
[6]Dipartimento di Scienze Fisiche, Informatiche, Matematiche (FIM), Università di Modena e Reggio Emilia, 41125 Modena, Italy
[7]Nanoscience Institute, National Research Council (CNR-NANO), 41125 Modena, Italy
[8]Department of Physics and Earth Sciences, University of Parma, 43124 Parma, Italy
[9]Center for Computational Chemistry, School of Chemistry, University of Bristol, BS8 1TS Bristol, UK
[†]these authors contributed equally to this work
[*]Corresponding authors: Carlo A. Pignedoli (carlo.pignedoli@empa.ch), Giovanni Pizzi (giovanni.pizzi@psi.ch)

## Workflow parameter settings panels

Although each workflow already encapsulates a wealth of expert knowledge, DFT simulations often involve fine-tuning numerous input numerical parameters, such as basis set size (e.g., wavefunction and charge-density cutoffs in the case of plane-wave methods as is the case for QUANTUM ESPRESSO), **k**-point sampling, choice of type and magnitude of the smearing of the electronic states, or convergence thresholds. The workflow configuration step aims to balance abstraction of complexities involved in calculation setup with flexibility of customizing calculation parameters often required by advanced users. To do so, the app splits the step into basic, advanced, and property-specific panels, each covering one aspect of this goal. Below, we describe briefly the features and purpose of each panel.

### *Basic settings - top-level controls and calculation protocols*

To streamline workflow setup, we provide in this panel top-level controls to specify general considerations for the system (metallic/insulating, magnetic, spin-orbit coupling - see Fig. S1). In addition, to abstract the complexities of configuring DFT calculations, we implement a protocol system which pre-configures a curated set of computational settings for different levels of accuracy and computational cost (Table S1). We provide three predefined protocols, each tailored to a specific use case and benchmarked in Ref. 1:

1. **Fast**: Optimized for speed, this protocol is intended only for testing purposes or quick, preliminary investigations where computational efficiency is the priority.

2. **Balanced**: An intermediate option that is typically recommended as the default, as it provides a balanced trade-off between computational cost and numerical precision.

3. **Stringent**: Designed for highly precise results, this protocol employs more stringent settings, suitable for cases where precision is paramount at the expense of a higher computational cost.

Other than the parameters shown in Table S1, the protocols also affect other workflow-level settings, such as the volume convergence used in structural relaxations. In addition, the "Stringent" protocol also selects a pseudopotential with higher plane-wave and charge-density cutoffs. All defaults can be overridden by users.

These protocols have been developed and tested for a large database of structures. By automatically setting key parameters like **k**-point meshes and convergence thresholds, the protocol system relieves users of the need for manual configuration, thereby reducing the potential for misconfiguration and improving overall workflow reliability.

**Table S1.** Protocol-dependent default parameters for self-consistent field (SCF) calculations.

| Protocol | Thresholds | | | Smearing width (Ry) | k-spacing ($\text{Å}^{-1}$) |
|---|---|---|---|---|---|
| | $E_{\text{SCF}}$ (Ry/atom) | $E_{\text{ionic}}$ (Ry/atom) | $F_{\text{ionic}}$ (Ry/bohr) | | |
| Fast | $4 \cdot 10^{-10}$ | $1 \cdot 10^{-4}$ | $4 \cdot 10^{-3}$ | 0.0275 | 0.3 |
| Balanced | $2 \cdot 10^{-10}$ | $1 \cdot 10^{-5}$ | $4 \cdot 10^{-4}$ | 0.02 | 0.15 |
| Stringent | $1 \cdot 10^{-10}$ | $5 \cdot 10^{-6}$ | $5 \cdot 10^{-5}$ | 0.0125 | 0.1 |

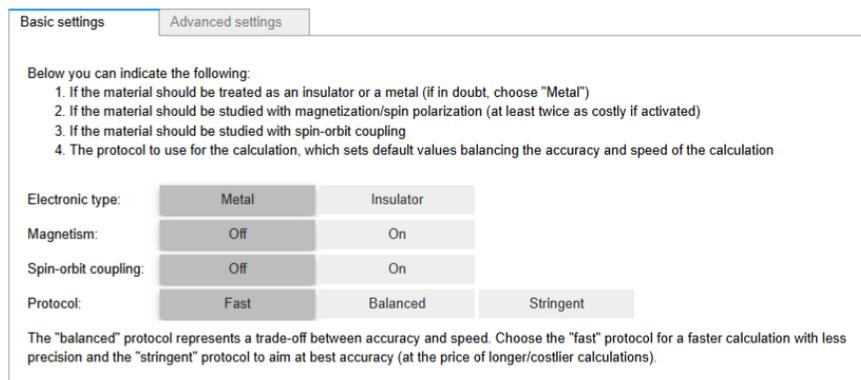

**Figure S1.** The basic parameter-settings panel providing users with top-level controls to streamline the process of configuring the calculation. These include defining the electronic type (metallic or insulating), switching on magnetism (spin polarization) and spin-orbit coupling, and choosing a calculation protocol, pre-configuring a set of calculation parameters balancing calculation speed and accuracy (see Table S1).

*Advanced settings - flexibility and tunability*
In this panel, we provide controls for users more familiar with the underlying code and theory to override protocol-specific defaults (convergence criteria, Brillouin zone **k**-point sampling, smearing, cutoffs) and customize system/site properties (magnetization, Hubbard parameters) and selected pseudopotentials (Fig. S2). The panel includes a reset button to re-apply protocol-dependent defaults. We note that for Hubbard parameters, though a plugin is available to compute Hubbard $U+V$ (see **Hubbard-parameter calculations** section below), we presently support in the app the use of Hubbard $U$ only. Development is ongoing to introduce controls for providing also Hubbard $V$.

*Plugin-specific settings*
Additional settings panels are provided for each selected plugin property, in which plugin developers provide users with a set of controls for tuning the calculation(s) involved in the plugin's workflows. Further details on plugin settings panels is given in the **Plugins** section below.

**Figure S2.** The advanced settings panel providing users with a wide range of controls to set convergence criteria, smearing, the $k$-point grid, Hubbard $U$ parameters, pseudopotentials, and wavefunction and charge density cutoffs. If magnetism is switched on (spin-polarized calculation), additional controls are displayed to parameterize the magnetic calculation (not shown).

## Plugins

The QE app supports running multiple properties (bands, PDOS, etc.) calculations in one app. The individual properties can be developed and seamlessly integrated into the app as plugins. This integration is made possible since for most properties, the following conditions are met:

- the configuration for a property calculation has its settings unrelated to other properties;
- the sub-workflow of the properties can be run independently;
- the analysis of the results of the properties is independent.

Each plugin is responsible for one property calculation. A plugin typically registers new panels (setting, result) and workflows (AiiDA WorkChains) to extend the functionality of the app. The plugin design makes the QE app modularized and pluggable. Consequently, developers have the flexibility to manage their plugins in a distinct folder within the QE app codebase, or they may choose to maintain it as an independent package.

### Muon spectroscopy

Muon spin rotation and relaxation spectroscopy is a fundamental experimental technique used to probe local magnetism, superconductivity and charge ordering mechanisms of materials[2]. The experimental setup is briefly summarized in the following. A positive muon with kinetic energy $E_k > 4$ MeV is implanted in a sample, losing gradually energy up to thermalization in a given (unknown) interstitial site in the crystal. The muon then interacts with the local magnetic

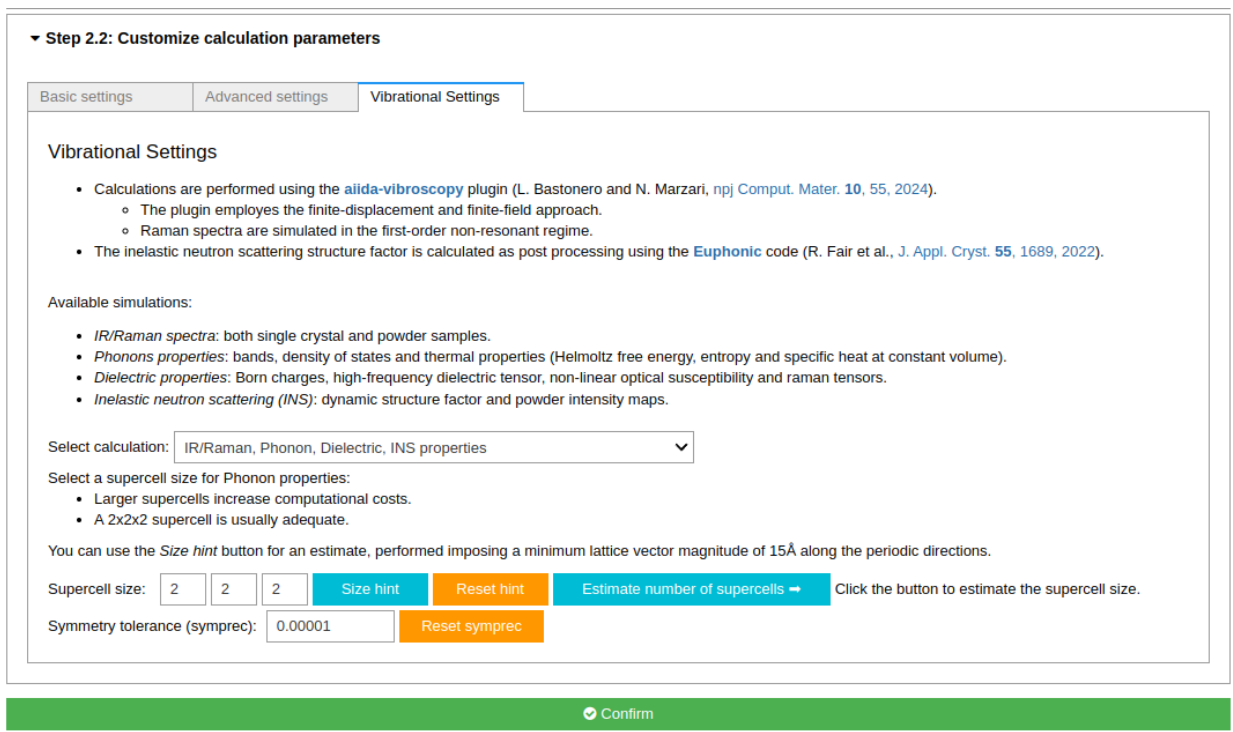

**Figure S3.** The configurations settings panel for the aiidalab-qe-vibroscopy plugin.

environment and, consequently, its spin polarization $P(t)$ undergoes a precession. The time evolution of $P(t)$ is the main experimental outcome. To correctly interpret the signal, it is fundamental to determine the muon resting sites in the material, as well as its local microscopic interactions. Experimentally, this is far from trivial and first-principles methods are therefore almost essential, as they can predict with great accuracy both the resting sites and the related local fields as felt by the muons. In particular, the approach called DFT+$\mu$[3,4] models the muon as an infinitely dilute impurity in the system and computes its property by means of DFT. This approach requires running a large number of supercell calculations, each with the muon placed in a different trial position. Once the muon resting sites are determined, it is possible to obtain the time evolution of the muon spin polarization originating from its interaction with nuclear magnetic dipoles in standard experimental conditions (i.e., when thermal energy is much larger than nuclear interactions), as e.g. computed by the UNDI package[5].

In the QE app, DFT+$\mu$ workflows are accessible by installing the `aiidalab-qe-muon` plugin[6]. The plugin implements the AiiDA `ImplantMuonWorkChain` workflow for the search of muon resting sites within a given material. The `ImplantMuonWorkChain` manages various sub-WorkChains dedicated to muon spectroscopy, implemented in the `aiida-muon` and `aiida-impuritysupercellconv` AiiDA plugins[7–10]. The former specializes in locating resting sites and magnetic interactions by means of DFT+$\mu$, while the latter is designed to automatically determine the appropriate supercell size for muon simulations (or, in general, infinitely dilute H impurities). A dedicated *Settings* panel allows users to configure and customize muon-related DFT+$\mu$ parameters, namely: supercell size, muon charge state, density of muon trial positions, inclusion of Hubbard corrections and spin polarization. If only a fast estimation of the needed supercell size is desired, this can instead be obtained live through a designated *Hint* button, bypassing explicit submissions of AiiDA workflows and utilizing instead heuristic estimates. Furthermore, it is possible to enable the post-processing calculation of the muon polarization (via the `UNDI` package) for predefined combinations of external magnetic field values and sample orientations. A results panel collects and visually summarizes all simulations outputs, as illustrated in Fig. S4. Specifically, all detected candidate muon resting sites are listed in a table alongside their associated relative total energy, with the lowest-energy one serving as the reference. A plot of the hosting lattice distortions is presented as useful hint to identify self-trapped muons. The sites are also visualized in a 3D viewer and users can seamlessly switch between two viewing modes: show each individual site within the corresponding supercell, or wrap all identified sites within the unit cell. If requested during the configuration phase, a prediction of the

depolarization function for each site is also provided. Finally, the supercells for each of the detected resting sites and the associated data shown in the results table can be easily downloaded for further post-processing.

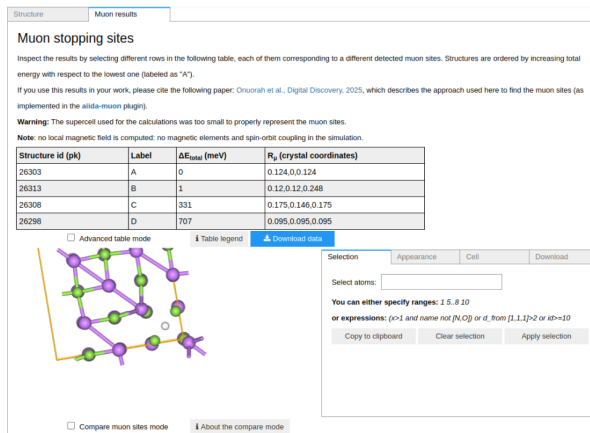

**Figure S4.** Results panel for the muon spectroscopy plugin, in the case of lithium fluoride LiF. The panel contains an informative table on the candidate muon sites and their relative total energy, as well as the position in the supercell, and possibly magnetic interactions (which are not shown in this case, as the system was not magnetic). Moreover, the supercell with the stopping muon site is presented. Users can interactively inspect it (for example computing the distance between the muon and the neighboring hosting atoms).

**X-ray core-level spectroscopies**

In the QE app, we provide additional plugins designed to automate the computation of core-level spectroscopies, covering X-ray photoelectron spectroscopy (XPS) and X-ray absorption spectroscopy (XAS) calculations.

XPS, a technique based on the photoelectric effect, stands as an important technique in modern surface and materials science. It finds extensive applications across diverse domains such as semiconductors, energy, environmental science, catalysis, and corrosion[11]. The so-called Delta Kohn-Sham (ΔKS) approach[12,13] is routinely used to rationalize XPS data. Starting from core-hole pseudopotentials, it allows one to compute the XPS binding energies as total energy differences of the systems with and without core hole for each given species in the target system[14]. We leverage the automated robust workflow to compute XPS spectra implemented in the aiida-qe-xpsec[15] package, which uses the ΔKS approach and pseudopotential-based total energy calculations, and we integrate it inside the QE app as a plugin, exposing it via a user-friendly interface. A dedicated configuration panel allows users to choose the system type (molecule or crystal), core-hole pseudopotential group, core levels, and target atoms for analysis (see Fig. S5). In the result panel, users can interactively adjust the spectral broadening and intensity parameters and view site-resolved shifts in a table that links directly to the 3D structural viewer (see Fig. S6). One can also upload reference experimental XPS data to compare with the calculated results.

Similarly, XAS has served for many decades an important role in materials science as a means to gain information about the local chemical environments present in a given material[16]. Theoretical approaches to analysing both the X-ray absorption near edge (XANES), up to 50 eV above the absorption onset, and the extended X-ray fine-structure (EXAFS) region beyond the near edge have been well established in the literature. Methods for computing XANES may vary by theoretical framework (for an overview see e.g. Refs. [17, 18]) and computational approach, including real-space approaches based on wavefunctions[19,20], multiple-scattering[21] as well as reciprocal-space, band-structure methods[22,23] within DFT. For the calculation of K-edge XANES, for which single-particle methods are known to describe with good accuracy[17], we provide a plugin exposing a simple and user-friendly interface to the underlying robust workflows implemented in the aiida-qe-xspec[15] package. These workflows use the XSpectra[24,25] code of QUANTUM ESPRESSO, adopting the projector augmented-wave (PAW) method in a reciprocal-space pseudopotential scheme, where the XANES spectrum is computed from a cross-section between a given core level (initial state) and the virtual (unoccupied) states (final states) of the material in the presence of a localized core-hole. The plugin can compute XANES spectra for any given solid exploiting crystal symmetry analysis to identify equivalent absorbing atom sites (if any). The choice of different core-hole treatments is offered to the user, i.e., full, half and excited core-hole[12,26], together with other options important to completely resolve the XANES spectrum of the material. Moreover, in the app

S5/S20

[Figure S5 screenshot: XPS configuration panel showing Basic settings / Advanced settings / XPS tabs, with Structure (Molecule/Crystal toggle, Crystal selected), Core-Hole pseudopotential group (Group: xps_pbe), Select core-level (C_1s, H not supported), and Select atoms (Indices field).]

**Figure S5.** Configuration panel for the XPS plugin. Users can define whether the system is a molecule or a crystal, select the pseudopotential group (including core-hole versions), choose specific core levels for analysis (e.g., C 1s), and optionally limit the calculation to specific atomic indices of the selected element.

we provide a set of well-tested core-hole pseudopotentials for several common elements (e.g. C, O, F). As an example, Figure S7 shows how the total K-edge for a selected element can be inspected in the XAS results panel, in addition to the K-edge XANES of sub-components belonging to symmetrically-inequivalent absorbing atoms. The plugin also features a set of post-processing tools to apply core-hole lifetime broadening effects to the computed spectrum based on the formulation of Bunau *et al.*[25] which closely matches the typical broadening seen in experimental spectra. The resulting spectra may then be downloaded for further analysis.

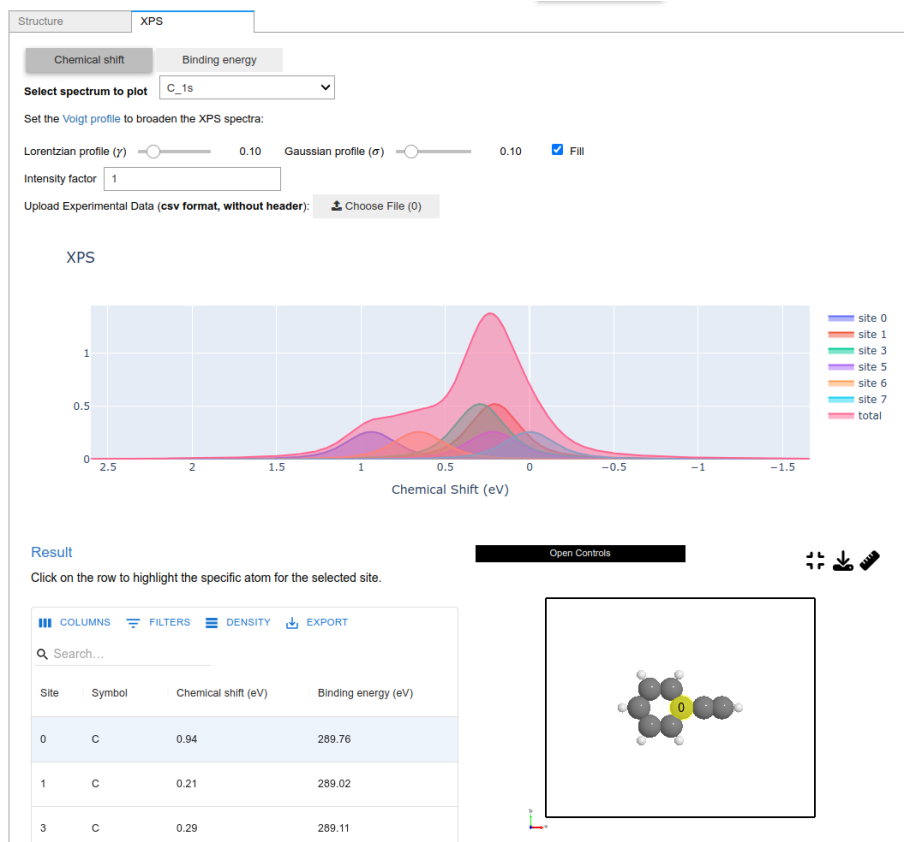

**Figure S6.** Results panel for the XPS plugin, showing the C 1s core level shift of phenylacetylene in gas phase. Users can interactively select the element (e.g., C), toggle between chemical shift and binding energy, and adjust the Voigt profile parameters (Lorentzian and Gaussian broadening) to simulate the experimental broadening effects. A summary table lists the chemical shift and binding energy values for each atomic site, with the option to highlight the corresponding atom in the 3D molecular structure upon selection.

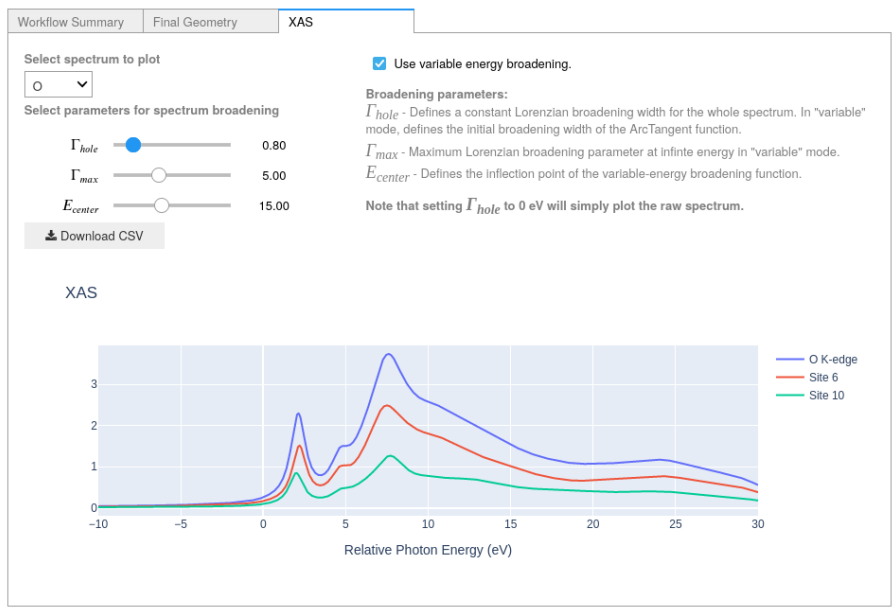

**Figure S7.** Results panel for the XAS plugin, showing the example of the O K-edge XANES of lithium carbonate ($Li_2CO_3$). Shown in the plot window are the XANES of the two symmetrically-inequivalent O sites of $Li_2CO_3$ (red/green) in addition to the full O K-edge XANES (blue). Users can interactively vary the spectrum broadening parameters (upper left of panel), switch between constant and variable broadening types (checkbox, top center of panel), and export spectra data in csv format (download button, center left of panel). Other edges calculated during the same workflow run may be selected using the dropdown menu in the top left of the panel.

**Wannier-function calculations**

The concept of maximally-localized Wannier functions (MLWFs) is important for understanding, interpreting and efficiently computing a wide range of electronic structure properties in condensed matter, from chemical bonding analyses to constructing tight-binding Hamiltonians for transport and topological characterizations[27–29]. Leveraging the open-source Wannier90 code[30], the plugin for Wannier functions (WFs) seamlessly orchestrates two main stages of the workflow (implemented in the `aiida-wannier90-workflows` package[31,32]) in an automated and user-friendly environment. A workflow is first launched to compute the SCF charge density and the corresponding band structure using DFT via QUANTUM ESPRESSO. The SCF density and bands are then used as inputs to a `Wannier90OptimizeWorkChain`. Within the plugin configuration panel, users can select among various approaches including the selected columns of the density matrix (SCDM) approach[33], and projectability disentanglement (PDWF method)[31], as well as choose a "frozen" energy window method to tailor the localization procedure to the system band structure[31]. An additional checkbox triggers the optional calculation of the Wannier functions in real space. In order to generate the Fermi surface and compute de Haas–van Alphen oscillation frequencies, the user needs to activate the respective checkboxes. The Fermi surface is computed at the last step of the Wannierization procedure, using the Wannier90 code. de Haas–van Alphen oscillation frequencies are computed using the open-source SKEAF code[34].

Once the calculations are finished, the plugin presents a comprehensive result page. It superimposes the DFT and Wannier-interpolated band structures in an interactive plot for comparison, as shown in Fig. S8, highlighting their band distance (see Ref. 31 for the definition). Below, a summary table displays key outputs of the Wannier90 code including the number of Wannier functions, the final WF spreads ($\Omega_{tot}$), their individual components ($\Omega_D, \Omega_I, \Omega_{OD}$) and the band distance. Additional panels offer line plots for monitoring the iterative convergence of the localization spreads, a detailed table of Wannier centers and spreads (both initial and final), and a 3D viewer for the structure. By selecting any row in the table, users can simultaneously visualize the isosurfaces of the selected Wannier function and highlight the corresponding atoms that are close to the Wannier center in the 3D structure viewer, as shown in Fig. S8. The interactive 3D visualization is powered by `weas-widget`[35], which is also utilized in the post-processing plugin (see next section on Post-processing plots). An additional 2D plot visualizes the de Haas–van Alphen oscillation frequencies. The tight-binding model as well as the Fermi surface files are available for download in the "Downloads" panel.

Overall, the Wannier function plugin within the QE app streamlines the entire workflow, from plane-wave band structure computations to MLWF localization, offering an intuitive, graphical interface for both setup and post-processing.

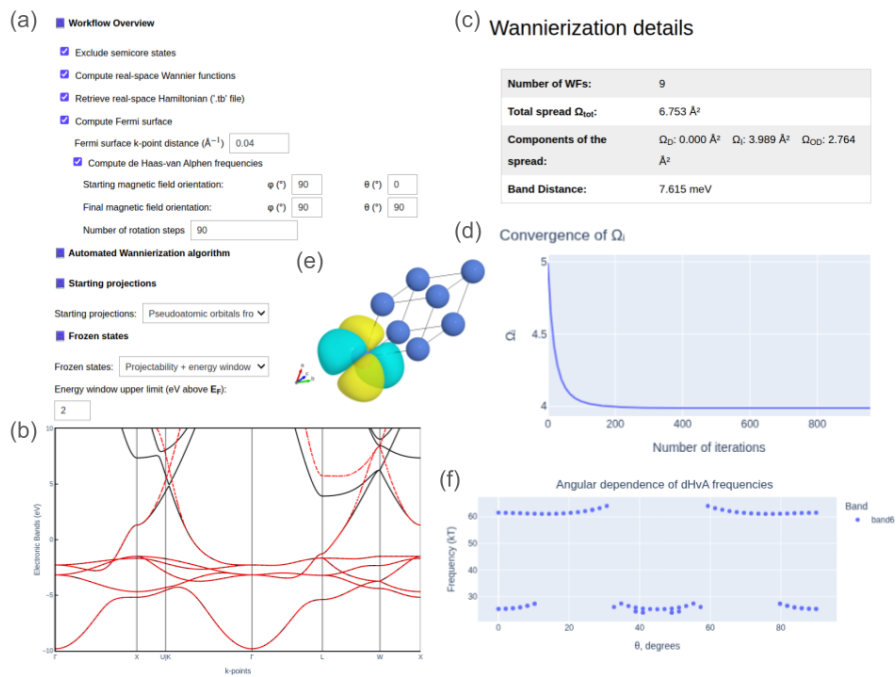

**Figure S8.** Panels for the Wannier function plugin in the QE app, illustrated in the case of a copper crystal. (a) Configuration panel used to set up the Wannierization workflow, including projection scheme, energy window, and optional generation of real-space Wannier functions, Fermi surface, and de Haas–van Alphen oscillation frequencies. (b) Comparison between the DFT band structure (black) and Wannier-interpolated bands (red), demonstrating the quality of interpolation. (c) Summary of Wannierization results showing the number of Wannier functions, spread components, and band distance between DFT and Wannier bands. (d) Convergence behavior of the Wannier function spread components. (e) 3D visualization of a selected maximally localized Wannier function, with its isosurface and nearby atoms highlighted in the structure viewer. (f) Angular dependence of de Haas–van Alphen oscillation frequencies.

**Hubbard-parameter calculations**

DFT with local or semilocal exchange-correlation functionals—such as the local-density approximation (LDA) and generalized-gradient approximation (GGA)—is a cornerstone of materials science. However, it suffers from significant self-interaction errors in systems with partially filled $d$- and/or $f$-shells, such as transition-metal and rare-earth compounds. To mitigate these errors, various advanced approaches have been developed, among which Hubbard-corrected DFT (DFT+$U$[36–38] and DFT+$U$+$V$[39]) stands out for its balance between accuracy and computational efficiency. A key challenge of this approach is the determination of the Hubbard $U$ and $V$ parameters. While these are often tuned empirically using experimental data, first-principles methods offer a more reliable alternative. Among them, the supercell-based linear-response method[40] and its reformulation using DFPT in primitive cells[41,42] have proven to be both accurate and robust for computing Hubbard parameters.

The plugin provides an interface to the `aiida-hubbard` AiiDA plugin[43,44], based on the `hp.x` code[45] of QUANTUM ESPRESSO that uses DFPT to compute the on-site $U$ and inter-site Hubbard parameters $V$ in a self-consistent, fully *ab initio* manner, ensuring that both the electronic ground state and the Hubbard corrections converge in tandem. The plugin provides an intuitive panel for defining calculation modes, specifying which atoms (for $U$) or pairs of atoms (for $V$) require Hubbard corrections, and setting convergence tolerances (see Fig. S9a). Users begin by selecting either the one-shot or self-consistent method from a dropdown menu. Specifically, the one-shot mode performs a single pass, fixing the geometry and extracting $U$ (and optionally $V$) parameters for the specified orbitals,

while the self-consistent mode applies an iterative routine, wherein the structure can be re-optimized at each step. The Hubbard parameters are updated until convergence criteria on $U$ and $V$ (and optionally the structure) are satisfied. Next, users select whether to include only on-site $U$ or both on-site $U$ and inter-site $V$ corrections. A scrollable table automatically lists each atomic kind in the structure, allowing one to toggle corrections for specific sites and input an initial guess for $U$. For inter-site interactions, pairs of atomic kinds are similarly exposed, with fields to activate and set an initial guess for $V$. Finally, advanced settings such as parallelization over individual atoms and **q** points for DPFT calculations can be enabled through checkboxes. These options leverage the inherent parallelization capabilities of the `hp.x` code[45] to significantly reduce walltime, particularly for large simulation cells or complex materials. Once the calculation is complete, the DFT+$U$+$V$ plugin presents the Hubbard parameters in an interactive table alongside the atomic structure (see Fig. S9b). This feature allows users to directly correlate computed $U$ and $V$ values with specific atoms or atom pairs selected in the table. The computed values can then be used as inputs for subsequent DFT+$U$+$V$ calculations (e.g., phonons, IR/Raman, XAS, XPS, ...) or other advanced electronic structure analyses (see **Advanced settings – flexibility and tunability** above for how to manually set Hubbard parameters in the app).

**Figure S9.** User interface of the plugin for calculating Hubbard parameters. (a) Setup panel for defining the Hubbard parameter calculation: users can select between one-shot and self-consistent modes, choose the calculation type (DFT+$U$ or DFT+$U$+$V$), and configure which atomic sites or atom pairs should receive on-site ($U$) or inter-site ($V$) corrections. Additional fields allow setting initial guesses and enabling parallelization over atoms and **q** points. (b) Results panel showing computed Hubbard parameters for selected atom pairs in a LiCoO$_2$ structure.

### Post-processing plugin

The QE app supports an external plugin to run post-processing calculations using `pp.x`, based on wavefunction files produced from previously executed PDOS or band structure calculations. To use the plugin, users must select a structure associated with a completed PDOS or Bands calculation. Fig. S10 shows the configuration panel of the plugin. It lists all relevant `PwCalculation` entries linked to the selected structure and provides essential metadata, such as whether LSDA or spin-orbit coupling was used. This information is required to determine the appropriate `pp.x` post-processing

logic. Additionally, the plugin displays the AiiDA computer on which the original calculation was performed. This is critical, as the post-processing must be executed on the same machine where the original calculation was run in order to access the needed files. Once the selected post-processing calculations are completed, the results will be displayed in a separate tab featuring a 3D plot, as shown in Fig. S11.

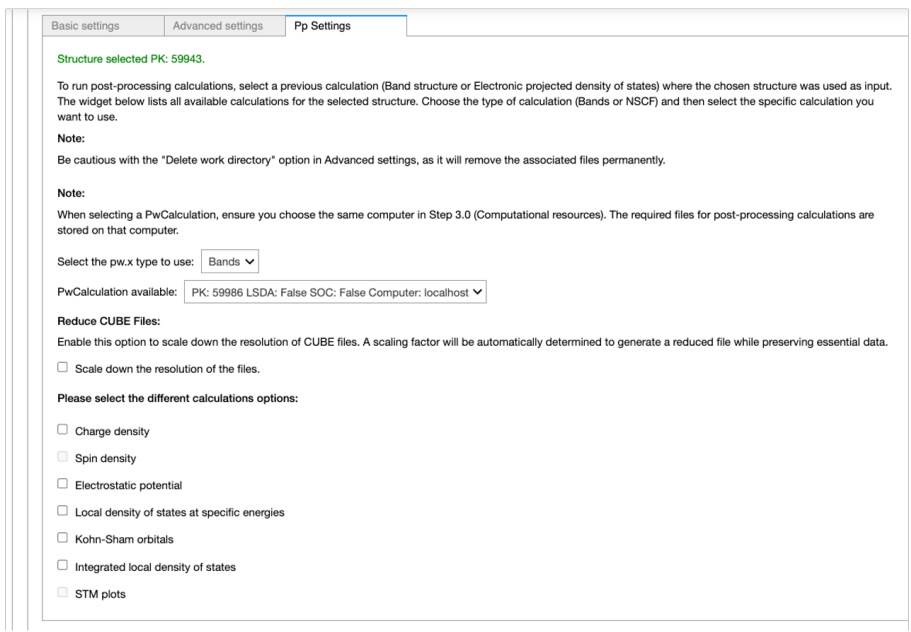

**Figure S10.** The configurations settings panel for the aiidalab-qe-pp plugin.

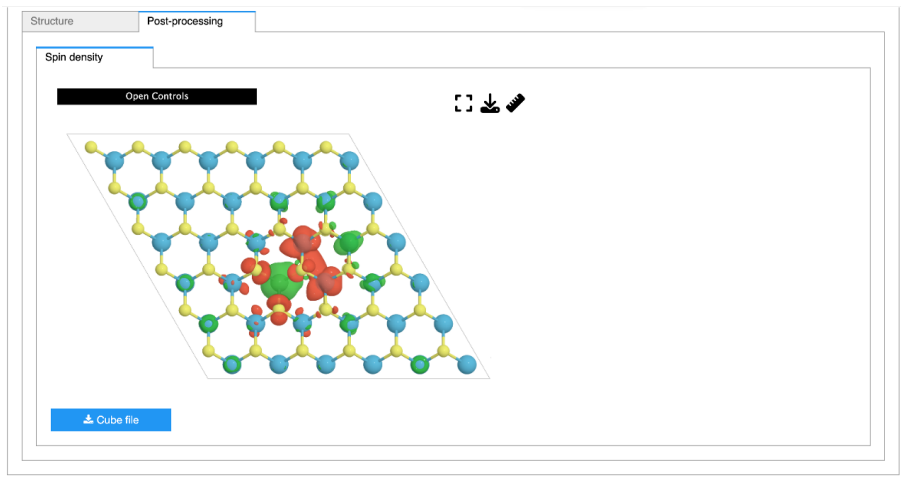

**Figure S11.** The result panels for the aiidalab-qe-pp plugin.

### Bader charge analysis
Bader analysis partitions the total charge density into zero-flux atomic basins, providing chemically intuitive net charges that are invaluable for quantifying charge transfer, oxidation states, and catalytic activity[46]. The Bader plugin displays the per-atom charges in an interactive table alongside the three-dimensional structural view, as shown in Fig. S12.

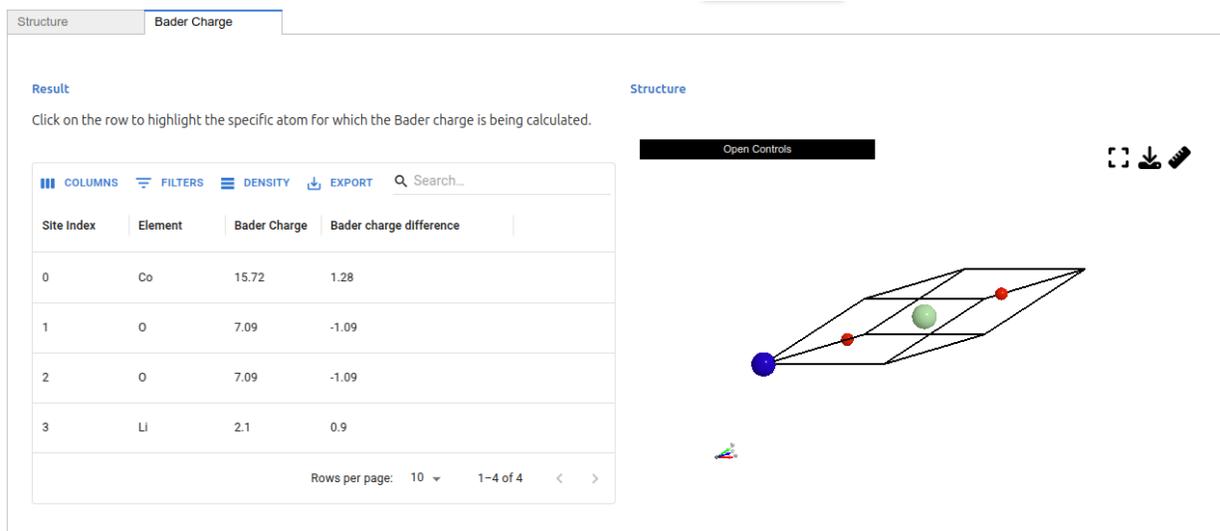

**Figure S12.** The result panels for the Bader charge plugin.

**In-app guides**

To streamline the use of the app by domain experts unfamiliar with the underlying simulation codes, we developed a system of internal guides/tutorials (Fig. S13). On selection, the activated guide injects additional information on adjacent widgets and controls, as well as a set of actions for users to take to complete the relevant steps involved in building, submitting, and visualizing results of the workflow pertaining to the subject of the guide. The guide system was developed with the plugin infrastructure in mind, allowing plugin developers to latch onto existing guide hooks, as well as introduce additional hooks within their plugin-specific components to provide description and tasks focused on the plugin's features (Fig. S13d-e). Encouraging further exploration of the app, we also provide post-guide exercises for users to try, including external resources for further learning (Fig. S13f).

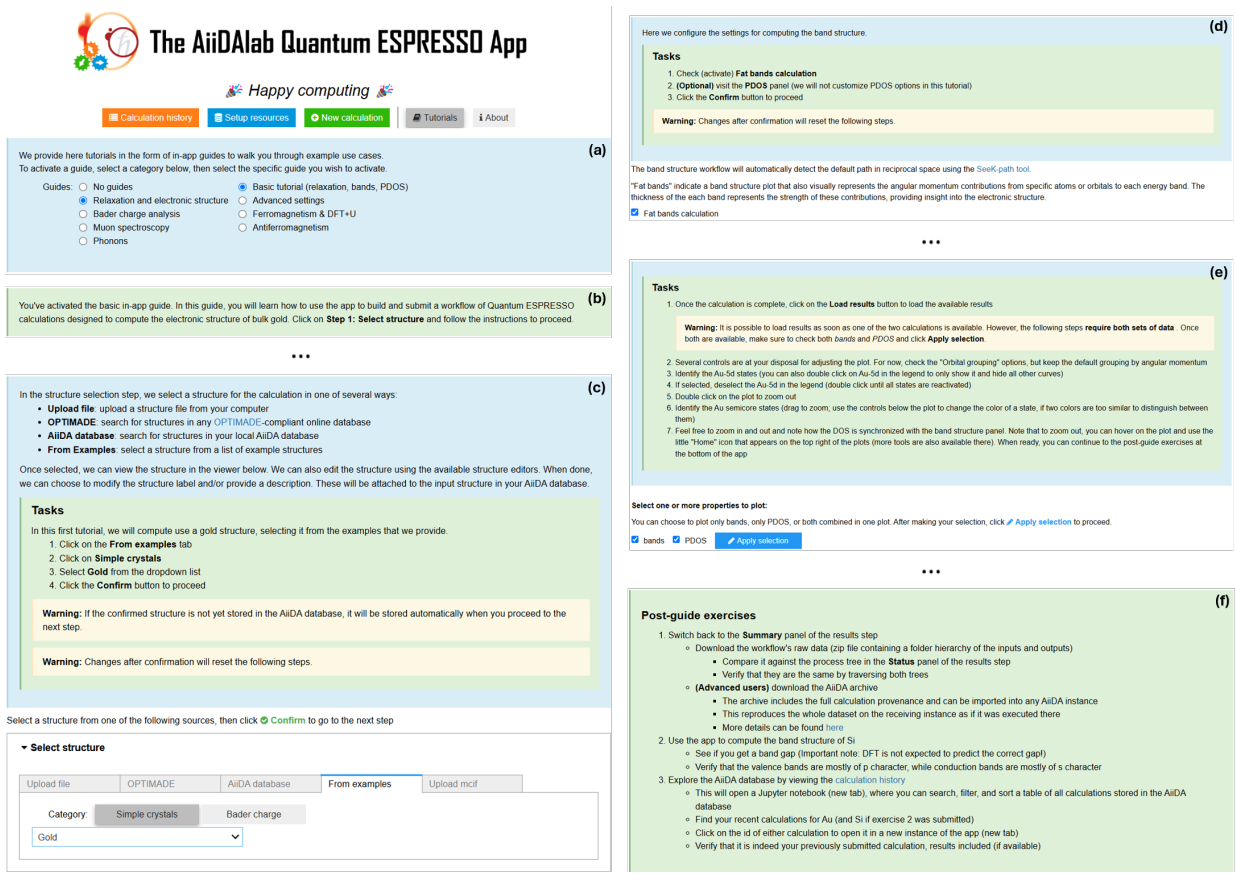

**Figure S13.** The QE app's in-app guides, here showing (a) the guide selection panel, (b) the guide banner, (c) a built-in structure panel guide, electronic structure plugin guides for (d) parameter settings and (e) results analysis, and (f) post-guide wrap up and further exploration.

## Utilities

### Calculation history

To facilitate better job management, the QE app provides a calculation history interface that allows users to efficiently search and manage their computational jobs run by the app. Fig. S14 illustrates the calculation history page. The page displays a table of all jobs in the database, along with various filtering options to help finding specific jobs. The table includes columns for the job ID, structure, creation time, state, label, relax type, and properties associated with each job. Additionally, each row in the table provides links to delete the respective job or to download its files. Users can filter the displayed data using various filters: properties, job state, label search, and date range. On the delete page, the details of a specific job are shown. Before deletion, the system checks for any dependent calculations linked to the node. If dependencies are found, the deletion is halted, and a warning is displayed. If no dependencies are present, the system asks for explicit user confirmation to ensure the action irreversibility is understood, preventing accidental deletion.

**Figure S14.** Calculation history interface of the QE app. The interface displays a searchable and filterable table of all previously run jobs, showing essential metadata such as job ID, structure, creation time, state, and associated properties. Users can edit job labels, delete or download results, and apply filters based on job type, date range, and computed properties. The interface aids in managing and tracking simulation workflows efficiently.

**Plugin management**

The QE app also includes a plugin management page that simplifies how users find and manage their plugins, making it easier to customize and enhance their computational environment. Users can browse and install plugins available in the official AiiDAlab Quantum ESPRESSO Plugin registry directly from the app. Developers are encouraged to register their plugins in this registry, which includes essential information such as the plugin's name, a link to its Git repository, documentation, and a categorization of its functions.

Figure S15 showcases the plugin management interface in the app. Once a plugin is integrated into the registry, it is automatically listed on the app's management page, making it accessible to every AiiDAlab user. This infrastructure not only aids developers in sharing their code and expertise with the community, but also allows them to maintain complete control over the development process of their GUIs.

**Figure S15.** Plugin management interface of the QE app. The interface provides a centralized view of all available Quantum ESPRESSO plugins from the official registry, allowing users to browse, install, or remove plugins with a single click. Each plugin entry includes metadata such as name, description, author(s), links to documentation and source code, and status indicators for installed plugins. This system supports modular extension of the app, enabling users to tailor their simulation environment and developers to distribute and maintain their plugins efficiently.

**Computational resources setup**

Though not an integrated part of the app, we maintain an external notebook as part of the AiiDAlab ecosystem dedicated for managing computational resources (Fig. S16). The notebook allows users to create AiiDA codes, which reference executables on local (AiiDAlab container) or remote machines. For remote machines, the notebook provides a set of widgets for establishing parameters for the SSH connection, defining the machine itself, and specifying the code executable on the machine. The widgets represent a one-to-one UI to AiiDA's resource creation API. The notebook also provides a table of available codes on the container. Users can browse through the table (using filters for convenience) and optionally hide codes to exclude them from code selectors in the app (in step 3).

**Figure S16.** External resource management notebook allowing users to define AiiDA references to remote executables. On the left, users can select pre-defined remote machines and select from a set of common QE codes. Advanced users more familiar with AiiDA can always opt for full control over the process by ticking the provided checkbox. On the right, a list of available codes in the current AiiDA instance is provided to users. Codes may be marked as hidden, thus excluding them from code selectors in the app (step 3).

## App examples

We provide a `Download Examples` page to help users quickly understand the capabilities of the app and its available plugins. Users can download predefined examples (such as electronic band structures, phonons, muons spectroscopies, and more), load them directly into the app, and explore the complete workflow. By inspecting each step, especially the Results Panel, users gain a clear and comprehensive understanding of what the app has to offer. This allows them to quickly assess whether the tool meets their specific needs. As shown in Fig. S17, users land on a page where they can browse a descriptive list of available examples.

Multiple examples can be selected at once. By clicking the Import button, the chosen examples are automatically downloaded and imported into the AiiDA database. Once imported, the examples appear in the Calculation History (see Fig. S14) each with a recognizable label that corresponds to the example list. From there, users can open and inspect each example in detail.

**Figure S17.** The *Download Examples* page allows users to select one or more predefined examples from a descriptive *Examples* list. By clicking the "Import" button, the selected examples are automatically downloaded and imported into the AiiDA database. They are then available in the Calculation History with recognizable labels, ready to be opened and explored.

### Model-View-Controller

The QE app follows a standard **Model-View-Controller** (MVC) design patten, where the traits of **View-Controller** widgets - here referring to `traitlets` traits - are maintained in `Model` subclasses, where `Model` is a simple extension of the `traitlets.HasTraits` base class to provide `Mixin` support for ease of further extension. The different components of the app are connected via a network of models, leveraging both the **Observer** and **Moderator** design patterns to facilitate data transfer by using the trait observation and linking functionalities provided in the `traitlets` library.

Changes in the state of a given model propagate through the model network, triggering subscribed observables to update local states that are observed by the user interface. This design sets up clear communication between user interactions and the app's data model, from workflow setup, to submission, to monitoring, and analyzing of results.

To simplify development and maintenance, the model network is constructed with a set of dedicated trait-aware `Mixin` classes, providing both specific traits and functionality, for example the `HasInputStructure` mixin with its `input_structure` trait and `has_pbc` property method, or the `HasProcess` mixin providing a `process_uuid` and a host of dedicated methods to interact with AiiDA processes.

The separation of responsibilities provided by the MVC design pattern simplifies system and integration testing by dedicating separate testing suites for UI and models. More importantly, decoupling the UI from the data model provides clear patterns for development of the front-end and more strictly defines the app's source of truth in the model network.

Lastly, the separation of the UI and the data model enables the implementation of lazy-loading techniques and mechanisms in the UI, ensuring that only those widgets that are required to be loaded at a given moment are loaded, with others only loaded on user demand, usually by means of opening an accordion panel, or switching to an unvisited tab. Such built-in `ipywidgets` container widgets have been equipped with loading (spinner) widgets via a generic

`LoadingWidget` class to improve communication of UI loading to users. Overall, the implementation of lazy-loading techniques has significantly improved the loading time of the app by deferring/distributing minor loading times to each widget.


\endgroup
\clearpage
\end{document}